\documentclass[twocolumn]{aastex62}

\received{}
\revised{}
\accepted{}
\submitjournal{ApJ}

\usepackage{placeins}
\usepackage{amsmath}
\usepackage{color}
\usepackage{enumerate}

\usepackage{graphicx}
\usepackage{tabularx,booktabs}
\usepackage{grffile}
\usepackage[varg]{txfonts}
\usepackage{subfigure}
\usepackage{cancel}
\usepackage{bm}
\usepackage{hyperref}

\begin{document}

\newcommand{\rev}[1]{{\bf #1}}

\newcommand\msun{$M_\odot$}
\newcommand\lsun{$L_\odot$}
\newcommand\methanol{CH$_3$OH}
\newcommand\hcop{HCO$^+$}

\shorttitle{The chemical inventory of the planet-hosting disk PDS~70} 
\shortauthors{Facchini et al.}

\title{The chemical inventory of the planet-hosting disk PDS~70}

\correspondingauthor{Stefano Facchini}
\email{stefano.facchini@eso.org}

\author[0000-0003-4689-2684]{Stefano Facchini}
\affiliation{European Southern Observatory, Karl-Schwarzschild-Str. 2, 85748 Garching, Germany}

\author[0000-0003-1534-5186]{Richard Teague}
\affiliation{Center for Astrophysics $|$ Harvard \& Smithsonian, 60 Garden Street, Cambridge, MA 02138, USA}

\author[0000-0001-7258-770X]{Jaehan Bae}
\altaffiliation{NASA Hubble Fellowship Program Sagan Fellow}
\affil{Earth and Planets Laboratory, Carnegie Institution for Science, 5241 Broad Branch Road NW, Washington, DC 20015, USA}

\author[0000-0002-7695-7605]{Myriam Benisty}
\affiliation{Unidad Mixta Internacional Franco-Chilena de Astronom\'ia, CNRS, UMI 3386. Departamento de Astronom\'ia, Universidad de Chile, Camino El Observatorio 1515, Las Condes, Santiago, Chile}
\affiliation{Univ. Grenoble Alpes, CNRS, IPAG, 38000 Grenoble, France}

\author[0000-0001-7250-074X]{Miriam Keppler}
\affiliation{Max Planck Institute for Astronomy, K\"onigstuhl 17, 69117, Heidelberg, Germany}

\author[0000-0002-0786-7307]{Andrea Isella}
\affiliation{Department of Physics and Astronomy, Rice University, 
6100 Main Street, MS-108,
Houston, TX 77005, USA}

\begin{abstract}

As host to two accreting planets, PDS 70 provides a unique opportunity to probe the chemical complexity of atmosphere-forming material. We present ALMA Band 6 observations of the PDS~70 disk and report the first chemical inventory of the system. With a spatial resolution of $0\farcs4-0\farcs5$ ( $\sim50\,$au), 12 species are detected, including CO isotopologues and formaldehyde, small hydrocarbons, HCN and HCO$^+$ isotopologues, and S-bearing molecules. SO and CH$_3$OH are not detected. All lines show a large cavity at the center of the disk, indicative of the deep gap carved by the massive planets. The radial profiles of the line emission are compared to the (sub-)mm continuum and infrared scattered light intensity profiles. Different molecular transitions peak at different radii, revealing the complex interplay between density, temperature and chemistry in setting molecular abundances. Column densities and optical depth profiles are derived for all detected molecules, and upper limits obtained for the non detections. Excitation temperature is obtained for H$_2$CO. Deuteration and nitrogen fractionation profiles from the hydro-cyanide lines show radially increasing fractionation levels. Comparison of the disk chemical inventory to grids of chemical models from the literature strongly suggests a disk molecular layer hosting a carbon to oxygen ratio C/O$>$1, thus providing for the first time compelling evidence of planets actively accreting high C/O ratio gas at present time. 

\end{abstract}


\section{Introduction}
\label{sec:introduction}

The frequency and characteristics of exoplanets detected in the last two decades indicate that planet formation is a robust and efficient mechanism that can lead to an outstanding diversity in the orbital architectures of planet systems and the physical properties of the planets within them  \citep{Winn2015}. The properties and architecture of exoplanetary systems are likely inherited from the physical and chemical processes that led to their formation in their birth environment, the protoplanetary disks. Key properties such as mass, radius, orbital period, multiplicity are likely determined by the formation mechanisms and subsequent evolution of  planetary systems \citep{Mordasini2018}. Possible signatures of forming planets, and of planet-disk interactions, are routinely detected as substructures in high resolution continuum observations of protoplanetary disks \citep{Andrews2020}, and as local deviations from Keplerian rotation in the gas velocity field \citep{DDC2020}. The tremendous advancement in observations of protoplanetary disks can now allow us to link the properties of the mature exoplanet population with the physical and chemical properties of their natal environments.

The chemical composition of planetary atmospheres, in particular, is thought to hold an imprint of the planet formation history, as the elemental composition directly traces the elemental abundances that have been accreted onto the planet surface through its formation phases \citep{oberg2011, oberg2016}. Atmospheric retrieval techniques have been used to determine the molecular and atomic abundances of a few species of massive exoplanets. With the aid of sophisticated chemical models, transmission and thermal emission spectra of super-Jovian planets allow an inference of the elemental ratio present in their atmosphere, in particular the C/O ratio, which is critical in determining the chemical processes that occur within it \citep[e.g.,][]{madhu2019,molliere_ea_2020}. At the same time, molecular line observations of protoplanetary disks have been used to derive volatile C/O ratios. C and O abundances in the warm molecular layer have been determined in a handful of objects, showing that elements are reprocessed between ice and volatile phases through physical and chemical processes \citep{bergin2016, Cleeves2018}.

The presence of massive planets embedded in their natal protoplanetary disks directly affects the physical and chemical properties of the material that they can accrete. On one hand, giant planets induce surface density and opacity variations across the disk, which change the thermal and ionization structure of the disk  \citep[e.g.,][]{Facchini2018,isella_and_turner_2018, Favre2019}. On the other hand, protoplanetary disks evolve both physically and chemically \citep[e.g.,][]{eistrup_ea_2018}, with a complex evolutionary path that is bound to determine the chemical abundances of gas and dust material that is accreted by embedded planets. In order to establish the chemical properties of material being accreted by massive planets, high angular resolution observations probing the chemical abundances of the hosting disks in the proximity of planets are needed.

This paper presents the first chemical inventory of the only disk known to harbor massive protoplanets, PDS~70, and as such lays a foundation stone in establishing the chemico-physical connection between massive planets and their host environment. The $\sim$5\,Myr old young star PDS~70 hosts two giant planets, PDS~70b and PDS~70c, recently directly imaged at multiple wavelengths \citep[e.g.,][]{keppler18,mueller2018, Wagner2018, haffert19, Christiaens2019, Mesa2019}. The planets are still accreting material from their host disk, as indicated by their strong emission in the H$\alpha$ line \citep{Wagner2018,haffert19}. The two planets are in a dynamically stable configuration, near 2:1 mean motion resonance \citep{Bae2019,wang_ea_2021}, with PDS~70b following an eccentric orbit and PDS~70c in a nearly circular one \citep{wang_ea_2021}. Other physical properties of the planets such as their mass are still rather unconstrained due to the likely presence of circumplanetary dust \citep[e.g.,][]{isella19,Stolker2020} affecting their spectra. First attempts in characterizing their atmospheric properties are being performed, but their C/O ratio is still unconstrained \citep[][]{wang_ea_2021}. The planets are observed to sculpt the disk structure in both its gas and dust components. They are carving a large cavity in the dust traced by continuum mm emission, with a deep gap traced by CO emission at the orbital radius of the two planets \citep{Hashimoto12,Dong12,hashimoto15,keppler19}.

PDS~70 thus provides the best candidate where to establish a link between massive protoplanets and their host environment, since the disk gas and dust structure, planetary accretion rates, masses, orbital radii and dynamical stability have been determined observationally. In this paper, we present a spatially resolved chemical inventory of the disk around PDS~70 using observations obtained with the Atacama Large Millimeter Array (ALMA). We demonstrate how the embedded planets affect the radial dependence of the molecular properties of their hosting disk, while the disk chemistry is bound to determine the chemical abundances of the building planetary atmospheres.  In Section~\ref{sec:reduction}, we describe the observations and data reduction. In Section~\ref{sec:results}, we present the first observational results of the chemical survey, while in Section~\ref{sec:physics} we derive physical properties for the detected molecular transitions. Finally, in Section~\ref{sec:discussion} we discuss our findings and summarize our results in Section~\ref{sec:conclusions}.

\section{Observations and data reduction}

\label{sec:reduction}

This paper mostly presents results from ALMA Band 6 observations taken in March 2020 as part of the ALMA program \#2019.1.01619.S (PI S. Facchini). These data represent the short spacing of a program aiming at imaging molecular lines at $0\farcs1$ resolution in the system, for which the long baseline data have not been taken yet. A summary of observation dates and properties is reported in Tab.~\ref{tab:observation_dates}. Two spectral setups were used to target several molecular species (see Tab.~\ref{tab:spectral_setup}), with two execution blocks of $\sim40\,$min per spectral setup. In particular, the lower frequency setup in the 217.194 - 233.955 GHz spectral range consisted of 7 spectral windows, while the higher frequency setup had 9 spectral windows. In both cases one of the spectral windows has a bandwidth of $1.875\,$GHz to increase the continuum signal-to-noise ratio (snr) and to allow for good self-calibration of the data. These spectral windows are set to Frequency Division Mode (FDM) to allow for serendipitous line detection.  

\begin{deluxetable*}{lccccccc} 
	\tablecaption{Log of ALMA observations used in this paper. The spectral setup is specified in Tab.~\ref{tab:spectral_setup}. \label{tab:observation_dates}}
	\tablecolumns{8} 
	\tablewidth{\textwidth} 
	\tablehead{
		\colhead{Frequency range}                          &
		\colhead{Date}                       &
		\colhead{Antennas$^a$}                     & 
		\colhead{Baselines}                     & 		
        \colhead{Time on source}                  & 
        \colhead{PWV}            &
        \colhead{Bandpass/flux calibrator}           &
        \colhead{Phase calibrator}
		  }
\startdata
217.194 - 233.955 GHz	& 06 Mar 2020	& 43 &  15-783\,m & 42\,min & 2.2\,mm & J1427-4206 & J1407-4302\\
& 16 Mar 2020	& 46 &  15-969\,m & 42\,min & 2.7\,mm & J1427-4206 & J1407-4302\\
243.570 - 262.081 GHz	& 02 Mar 2020	& 44 &  15-783\,m & 38\,min & 1.6\,mm & J1427-4206 & J1407-4302\\
& 06 Mar 2020	& 44 &  15-783\,m & 38\,min & 2.4\,mm & J1427-4206 & J1407-4302\\
\enddata
\tablenotetext{}{$^{a}$ Number of antennas after flagging.}
\end{deluxetable*}

The data were pipeline calibrated using the \texttt{CASA} package, version 5.6. After minor manual flagging, we imaged the continuum of each execution block separately, and noticed that likely due to poor weather conditions the pointing was offset between execution blocks by a significant fraction of the beam. While this can corrected for with self-calibration, we preferred to re-center each execution block beforehand in order to construct the best model for self-calibration from the first step \citep{andrews18}. Since PDS~70 is a transition disk with a large cavity and that no clear emission from an inner disk is present in the single execution blocks, we were unable to align the individual execution blocks by fitting the peak continuum intensity at the disk center, as is often done. We thus fitted the outer ring using a Gaussian ring model in the visibility plane, with seven free parameters (intensity normalization, radius and width of the ring, inclination, position angle and disk center). Each execution was fitted separately using the \texttt{GALARIO} code \citep{galario}, sampling the posterior distributions with the \texttt{emcee} package \citep{Foreman-Mackey+13}, a Python implementation of the Metropolis-Hastings ensemble sampler, with 60 walkers and 2000 steps after burn-in. Flat priors were used, with the parameters initialized from the fit to the ALMA Band 7 continuum reported in \citet{keppler19}. Baselines were selected to be $<650\,$k$\lambda$ to have the most uniform $(u,v)$-coverage across different execution blocks. We then shifted the phase center of each execution block using the \texttt{fixvis} task, and re-aligned all the execution blocks to the same coordinate system using the \texttt{fixplanet} task.

Particular care was also given to the flux calibration of the observations. For all execution blocks, the Band 6 flux was corrected by deriving a spectral index from a Band 3 and Band 7 observation of the flux calibrator J1427-4207. While the Band 3 observation of the flux calibrator was always within 3 days from our science observations, the closest observation in Band 7 was from February 27, 2020, 18 days before the latest observation of PDS~70. We thus decided to correct for flux offsets between execution blocks with the same spectral setup by using the flux  from the observation that was closest in time to the Band 7 calibrator estimate. To do so, we followed the procedure used in the DSHARP collaboration \citep{andrews18}, correcting for flux offsets using the \texttt{gaincal} task. We emphasize that for the low frequency observations the closest flux calibrator measurement in Band 7 was 8 days apart from the science observation of the first execution block. 

\begin{table*}
\centering
\caption{Spectral windows of the two spectral settings of the ALMA Band 6 observations. Frequency and velocity resolutions are obtained after Hanning smoothing being performed directly at the correlator.}
\begin{tabularx}{0.73\textwidth}{cc ccl}
\toprule
Central rest & Frequency  & Velocity  & \#channels & Main lines included in spw \\
frequency (MHz) & resolution (kHz) & resolution (km\,s$^{-1}$) & & \\
\midrule
217238.530 & 244.141 & 0.389 & 480 & DCN $J$=3-2   \\
218222.192 & 244.141 & 0.388 & 480 & H$_2$CO 3$_{03}$-2$_{02}$  \\
218732.732 & 244.141 & 0.387 & 960 & H$_2$CO 3$_{21}$-2$_{20}$ \\  & & & & c-C$_3$H$_2$ 7$_{16}$-7$_{07}$ \\ 
219560.358 & 122.070 & 0.193 & 960 & C$^{18}$O $J$=2-1  \\
220398.684 & 122.070 & 0.192 & 960 & $^{13}$CO $J$=2-1  \\
230538.000 &  61.035 & 0.092 & 1920 & $^{12}$CO $J$=2-1 \\
233002.432 & 976.562 & 1.453 & 1920 & CH$_3$OH $10_2$ - $9_3$(A$^+$) \\
\midrule
244500.000 & 976.562 & 1.384   & 1920   & CS $J$=5-4 \\
 & & & & CH$_3$OH $5_1$ - $4_1$(A$^{-}$) \\
 & & & &  c-C$_3$H$_2$ 3$_{21}$-2$_{12}$ \\
246404.588 & 244.141 & 0.343   & 960    & SO 2$_3$-3$_2$ \\
246663.470 & 244.141 & 0.343   & 960    & $^{34}$SO 5$_6$-4$_5$ \\
258156.996 & 244.141 & 0.328   & 480    & HC$^{15}$N $J$=3-2 \\
258255.825 & 244.141 & 0.328   & 480    & SO 6$_6$-5$_5$ \\
259011.787 & 244.141 & 0.327   & 960    & H$^{13}$CN $J$=3-2 \\
260255.342 & 122.070 & 0.163   & 480    & H$^{13}$CO$^+$ $J$=3-2 \\
261843.721 & 122.070 & 0.162   & 480    & SO 7$_6$-6$_5$ \\
262004.260 & 122.070 & 0.161   & 960    & C$_2$H $N$=3-2, $J$=$\frac{7}{2}$-$\frac{5}{2}$, $F$=4-3 \\
 & & & & C$_2$H $N$=3-2, $J$=$\frac{7}{2}$-$\frac{5}{2}$, $F$=3-2 \\
 & & & & C$_2$H $N$=3-2, $J$=$\frac{5}{2}$-$\frac{3}{2}$, $F$=3-2 \\
\bottomrule
\end{tabularx}
\label{tab:spectral_setup}
\end{table*}

The datasets with same spectral setup were combined using the \texttt{concat} task, and then self-calibration was performed on the two frequency settings separately. All spectral windows were included in the self-calibration, after flagging channels within $\pm20\,$km\,s$^{-1}$ ($\sim14.5-17.5\,$MHz depending on the central rest frequency) from all the lines listed in Tab.~\ref{tab:spectral_setup}. The resulting continuum was channel averaged with channels $<500\,$MHz to avoid bandwidth smearing. At every step of the self-calibration the model was imaged cleaning down to $5\sigma$ of the initial data, with a Briggs robust parameter of $0.5$ and an elliptical mask with a position angle (PA) of $160.4^\circ$, semi-major axis of 1$\farcs$7 and a ratio of major to minor axis given by an inclination of $51.7^\circ$ \citep{keppler19}. The rms is estimated over an elliptical annulus with the same PA and axes ratio between 2$\farcs$55 and 4$\farcs$25. We performed three rounds of phase self-calibration in both frequency setups. A first one accounting for polarization offsets with scan length as solution interval, and then two rounds combining polarizations and spectral windows with 120 and 60\,s solution intervals for the low frequency setting, and 120 and 30\,s solution intervals for the high frequency one. A final step of amplitude self-calibration at scan length was performed in both cases. The overall improvement was 119\% and 54\%, for a peak snr of 434 and 503, for the low and high frequency setups respectively. Image reconstruction was performed with the \texttt{tclean} task by using the \texttt{multiscale} deconvolver, cleaning down to the $1\sigma$ level. More details about the continuum imaging are reported in Section~\ref{sec:continuum}. The bandwidth-weighted mean continuum frequencies are 225.56\,GHz (1.33\,mm) and 252.80\,GHz (1.19\,mm).

The self-calibration tables were then applied to the unflagged, unaveraged visibilities. Continuum subtraction was performed using first order fitting with the \texttt{uvcontsub} task. Imaging of the spectral lines was performed with \texttt{tclean} with the \texttt{multiscale} deconvolver, allowing for point sources in the emission (particularly important for the spatially unresolved high velocity channels). In the cleaning process, we apply so-called Keplerian masks, i.e. exploit the rotation of the disk well fitted by a Keplerian model in the $^{12}$CO $J$=3-2 line to have a prior on where the molecular lines originate. We use the following geometrical parameters: M$_{\rm star}=0.875\,M_\odot$, $i=51.7^\circ$, PA$=160.4^\circ$ \citep{keppler19}. For bright optically thick lines ($^{12}$CO and HCO$^+$) we accounted for an elevated disk emission surface in creating the maps, using the best fit parameters by \citet{keppler19}: $z(r)[\arcsec]=0.33\times(r / 1\arcsec)^{0.76}$, with $z$ and $r$ being the vertical and radial cylindrical coordinates. The channel masks were created with the \texttt{keplerian\_mask.py} tool\footnote{\url{https://github.com/richteague/keplerian\_mask}}.

As a first attempt, all lines listed in Tab.\ref{tab:spectral_setup} were imaged between -6 and +17\,km\,s$^{-1}$ with both natural and Briggs weighting with a robust parameter equalling zero. Given the very different line intensities spanning two orders of magnitudes, multiple imaging runs have been performed on each line to obtain the best compromise between angular resolution, intensity and channel spacing. The choice of the imaging parameters for each line cube depends on the scientific question that is being addressed, e.g., natural weighting and uvtapering is used to maximize line detection, while Briggs weighting with robust parameter of 0 is used to maximize the spatial resolution to extract intensity profiles of bright lines. Tab.\ref{tab:fluxes} summarizes the fiducial imaging parameters chosen for the lines included in this study.

\begin{figure*}[t]
\begin{center}
\includegraphics[width=0.49\textwidth]{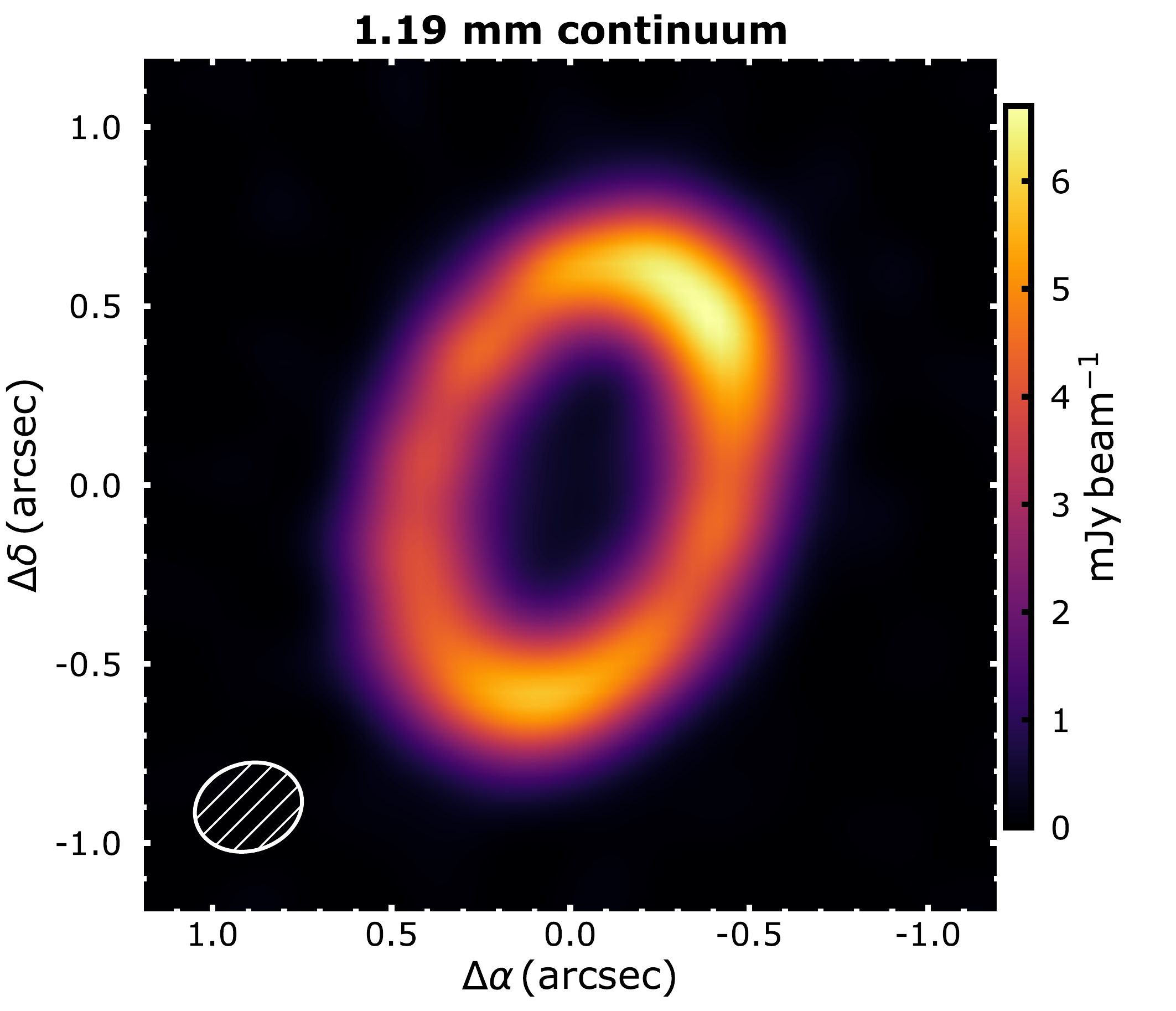}
\includegraphics[width=0.49\textwidth]{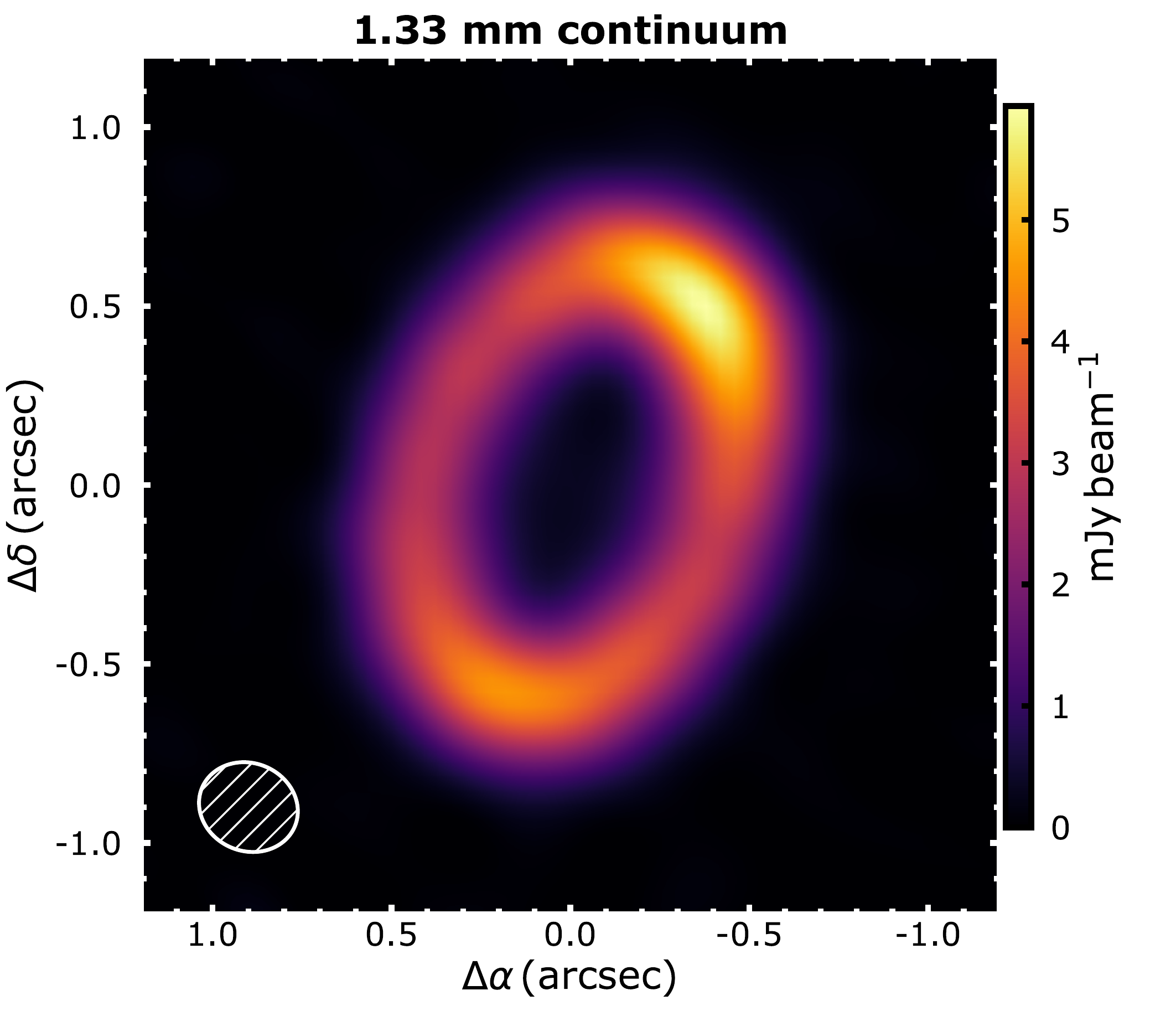}
\end{center}
\caption{Intensity maps of the continuum emission at 1.19\,mm (left) and 1.33\,mm (right), imaged with \texttt{superuniform} weighting.}
\label{fig:continuum}
\end{figure*}

Band 7 observations of HCO$^+$ $J$=4-3 and $^{12}$CO $J$=3-2 are included in the analysis. In particular, data from programs \#2015.1.00888.S and \#2017.A.00006.S include the $^{12}$CO transition, while program \#2015.1.00888.S is reduced and analyzed for HCO$^+$. The \#2015.1.00888.S and \#2017.A.00006.S data were calibrated as described in \citet{isella19}. The gain solutions from the self-calibration were applied to the lines, which were imaged following the same procedure described above. In both cases, the angular resolution was high enough that natural weighting is used throughout the whole paper for both lines. The properties of the channel maps are reported in Tab.~\ref{tab:fluxes}.

\section{Results}
\label{sec:results}

\subsection{Continuum}
\label{sec:continuum}

We image the continuum of both spectral setups. When cleaning with Briggs weighting with a robust parameter of 2, we obtain a very high peak snr of 714 and 608 for the 1.19\,mm and 1.33$\,$mm images, respectively. Within the same mask used for self-calibration the disk shows a flux density of 68.7 and 57.6\,mJy, respectively, with an rms of 21\,$\mu$Jy\,beam$^{-1}$ in both images. The flux density at 1.33\,mm is consistent with the Submillimeter Array (SMA) observations presented in \citet{hashimoto15}. While they report a flux density of $\sim38$\,mJy, a new calibration of the data leads to a flux density of 63\,mJy for baselines $<15\,$k$\lambda$ (Mark Gurwell, private communication). The offset in the $^{12}$CO $J$=2-1 flux between the SMA and our ALMA data can also be ascribed to the same calibration difference.

Given the high snr of the robust$=2$ images, we re-image the data at high angular resolution using \texttt{superuniform} weighting, with a \texttt{npixels} parameter equalling 3 in the \texttt{tclean} task. While rarely used, the \texttt{superuniform} weighting leads to high angular resolution, at the cost of reducing the maximum recoverable scale and resulting in PSF with strong side-lobes. There is no flux density loss in the imaging. The \texttt{superuniform} images are shown in Fig.~\ref{fig:continuum}. The 1.19\,mm and 1.33$\,$mm images respectively present a synthesized beam of 0$\farcs$30$\times$0$\farcs$24  (PA = 107.3$^\circ$) and 0$\farcs$28$\times$0$\farcs$24 (PA = 67.9$^\circ$), with an rms of 71 and 53\,$\mu$Jy\,beam$^{-1}$.

Both images show an intensity map that is similar to what has been reported by \citet{Long18,keppler19} at 0.85\,mm, with a well-resolved bright ring exhibiting an over-brightness in the North-West. We detect an inner disk in the 1.33\,mm image, even though the outer ring still contributes to the observed intensity at the center of the cavity due to the limited spatial resolution and high dynamic range. Different imaging parameters aimed at suppressing the side-lobes of the PSF (e.g. Briggs weighting with low robust values, or \texttt{uniform} weighting) do not lead to a detection of the inner disk since they result in lower angular resolution images. By estimating the intensity at the center of the disk, we determine an upper limit for the inner disk flux density of 0.34\,mJy at 1.33\,mm. 

We compare the flux density of the disk with the one estimated by \citet{isella19}, who re-analyzed the 0.85\,mm data by \citet{keppler19} correcting for a flux density offset. By using their estimated flux density of 177\,mJy, we obtain a spectral index of $2.49\pm0.44$ and $2.84\pm0.59$ in the 0.85-1.33 and 0.85-1.19\,mm range. The uncertainties are dominated by the 10\% absolute flux calibration in all datasets. The derived spectral index is in line with the spectral indices of transition disks being on the high side within the protoplanetary disk sample \citep{pinilla2014,ansdell2018,tazzari2020}. By using the upper limit on the inner disk flux density at 1.33\,mm, we obtain a lower limit on the 0.85-1.33\,mm spectral index of the inner disk of $1.61\pm0.44$, indicative that the central emission is indeed affected by the outer disk, which artificially increases the flux density at longer wavelengths, thus reducing the spectral index. Dust thermal emission at warm temperatures in the proximity of the star would imply spectral indices $\geq2$ \citep[e.g.,][]{testi_ea_2014}.

\begin{deluxetable*}{lcccccccc} 
	\tablecaption{List of the main molecular lines with associated imaging parameters and flux. \label{tab:fluxes}}
	\tablecolumns{8} 
	\tablewidth{\textwidth} 
	\tablehead{
		\colhead{Species}                          &
		\colhead{Transition}                       &
		\colhead{Frequency}                     & 
		\colhead{Weighting$^a$}                     & 		
        \colhead{Beam (PA)}                  & 
        \colhead{Channel}            &
        \colhead{Rms$^b$}           &
        \colhead{Flux}          &
        \colhead{Uncertainty} \\
		\colhead{}        & 
		\colhead{}        &
		\colhead{(GHz)}        &
		\colhead{}          &
	    \colhead{}          &
		\colhead{(km\,s$^{-1}$)}        &
		\colhead{(mJy\,beam$^{-1}$)}        &
		\colhead{(mJy\,km\,s$^{-1}$)}        &
		\colhead{(mJy\,km\,s$^{-1}$)}  
		  }

\startdata
$^{12}$CO & 3-2 & 345.7959 & nat & 0$\farcs$11$\times$0$\farcs$10  (90.3$^\circ$) &  0.42     &  1.1          & 12421 & 147 \\
$^{12}$CO  & 2-1 & 230.5380 & r=0 & 0$\farcs$37$\times$0$\farcs$33 (67.1$^\circ$) & 0.09      & 4.1           &  6109 & 26 \\
$^{13}$CO  & 2-1 & 220.3986 & r=0 & 0$\farcs$39$\times$0$\farcs$35 (69.6$^\circ$) & 0.19      & 3.0           &  1846 & 14 \\
C$^{18}$O  & 2-1 & 219.5603 & r=0 & 0$\farcs$39$\times$0$\farcs$35 (68.3$^\circ$) & 0.19      &  2.4          &  465 & 9 \\
H$_2$CO  & 3$_{03}$-2$_{02}$ & 218.2221 & r=0 & 0$\farcs$40$\times$0$\farcs$35 (69.0$^\circ$) &  0.39     &  1.5           &  710 & 7 \\
H$_2$CO  & 3$_{21}$-2$_{20}$ & 218.7600 & nat & 0$\farcs$57$\times$0$\farcs$52 (100.8$^\circ$) &  0.39     &  1.2          &   53 & 12 \\
CH$_3$OH  & $10_2$ - $9_3$(A$^+$) & 232.4185 & nat & 0$\farcs$53$\times$0$\farcs$50 (96.7$^\circ$) & 1.45      &  0.6          &  $<$32$^{c}$ & ... \\
CH$_3$OH  & $5_1$ - $4_1$(A$^{-}$) & 243.9157 & nat & 0$\farcs$60$\times$0$\farcs$49 (107.0$^\circ$) & 1.38      &  0.7          &  $<$62$^{c}$ & ... \\
C$_2$H  &  $J$=$\frac{7}{2}$-$\frac{5}{2}$ $F$=4-3 & 262.0042 & r=0 & 0$\farcs$39$\times$0$\farcs$32 (107.7$^\circ$) & 0.16      &      3.0      &  741 & 12 \\
C$_2$H  &  $J$=$\frac{7}{2}$-$\frac{5}{2}$ $F$=3-2 & 262.0064 & r=0 & 0$\farcs$39$\times$0$\farcs$32 (107.7$^\circ$) & 0.16      &      3.0      &  586 & 14 \\
C$_2$H  &  $J$=$\frac{5}{2}$-$\frac{3}{2}$ $F$=3-2 & 262.0648 & r=0 & 0$\farcs$39$\times$0$\farcs$32 (107.7$^\circ$) & 0.16      &      3.0      &  603$^d$ & 17 \\
c-C$_3$H$_2$  & 3$_{21}$-2$_{12}$ & 244.2221 & nat & 0$\farcs$59$\times$0$\farcs$49 (107.0$^\circ$) & 1.38      &       0.7     &  60 & 11 \\
c-C$_3$H$_2$  & 7$_{16}$-7$_{07}$ & 218.7327 & nat & 0$\farcs$57$\times$0$\farcs$52 (100.8$^\circ$) & 1.00      &  0.9          &  $<$29$^{c}$ & ... \\
H$^{13}$CN  & 3-2 & 259.0117 & r=0 & 0$\farcs$39$\times$0$\farcs$32 (107.1$^\circ$) & 0.33      &  1.9          &  326 & 17 \\
HC$^{15}$N  & 3-2 & 258.1571 & nat & 0$\farcs$56$\times$0$\farcs$46 (100.6$^\circ$) & 0.33      &  1.7          &  197 & 16 \\
DCN  & 3-2 & 217.2386 & r=0 & 0$\farcs$40$\times$0$\farcs$35 (70.8$^\circ$) &   0.39    &   1.7         &  209 & 11 \\
HCO$^+$  & 4-3 & 356.7342 & nat & 0$\farcs$21$\times$0$\farcs$18 (73.1$^\circ$) &  0.21     &   3.3         &  5199 & 122 \\
H$^{13}$CO$^+$  & 3-2 & 260.2553 & r=0 & 0$\farcs$39$\times$0$\farcs$32 (108.8$^\circ$) &  0.16     &   3.0         &  344 & 12 \\
CS  & 5-4 & 244.9355 & r=0 & 0$\farcs$42$\times$0$\farcs$34 (108.8$^\circ$) &  1.38     &  0.9          &  450 & 6 \\
SO  & 7$_6$-6$_5$ & 261.8436 & nat & 0$\farcs$56$\times$0$\farcs$44 (103.6$^\circ$) & 1.38      &  1.1          &  $<$27$^{c}$ & ... \\
SO  & 6$_6$-5$_5$ & 258.2558 & nat & 0$\farcs$56$\times$0$\farcs$47 (104.4$^\circ$) & 1.38      &  1.0          &  $<$52$^{c}$ & ... \\
\enddata
\tablenotetext{}{$^a$ Natural weighting and Briggs weighting with 
\texttt{robust}=0 have been used; $^b$ Rms in one velocity channel; $^c$ $3\sigma$ upper limit; $^d$ Since only half of the spectral line was covered by the spectral setup, only part of the disk is used in the total flux estimate, assuming the emission is azimuthally symmetric. The flux uncertainty has also been increased accordingly.}
\end{deluxetable*}

\begin{figure*}
\begin{center}
\includegraphics[width=0.49\textwidth]{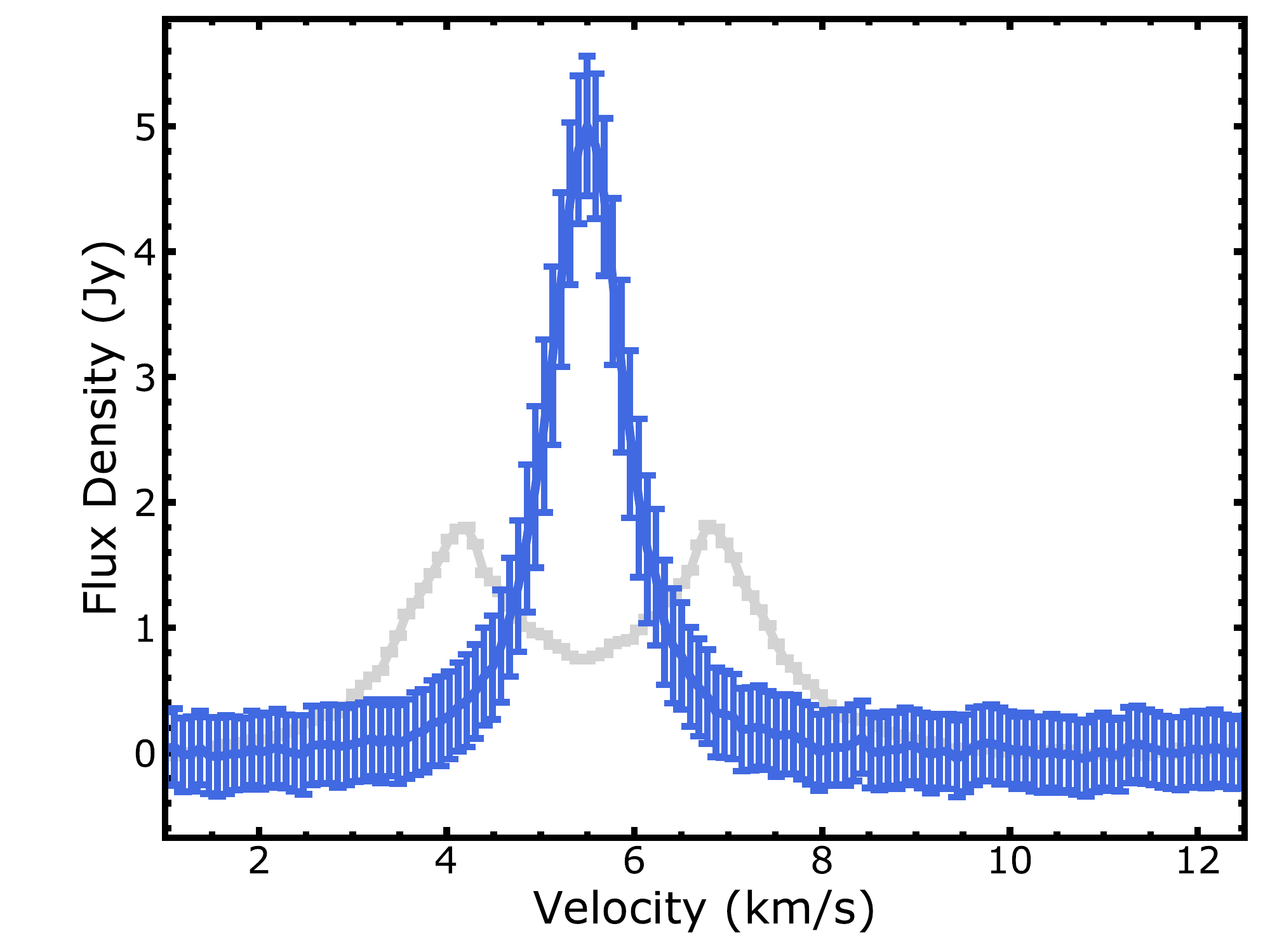}
\includegraphics[width=0.49\textwidth]{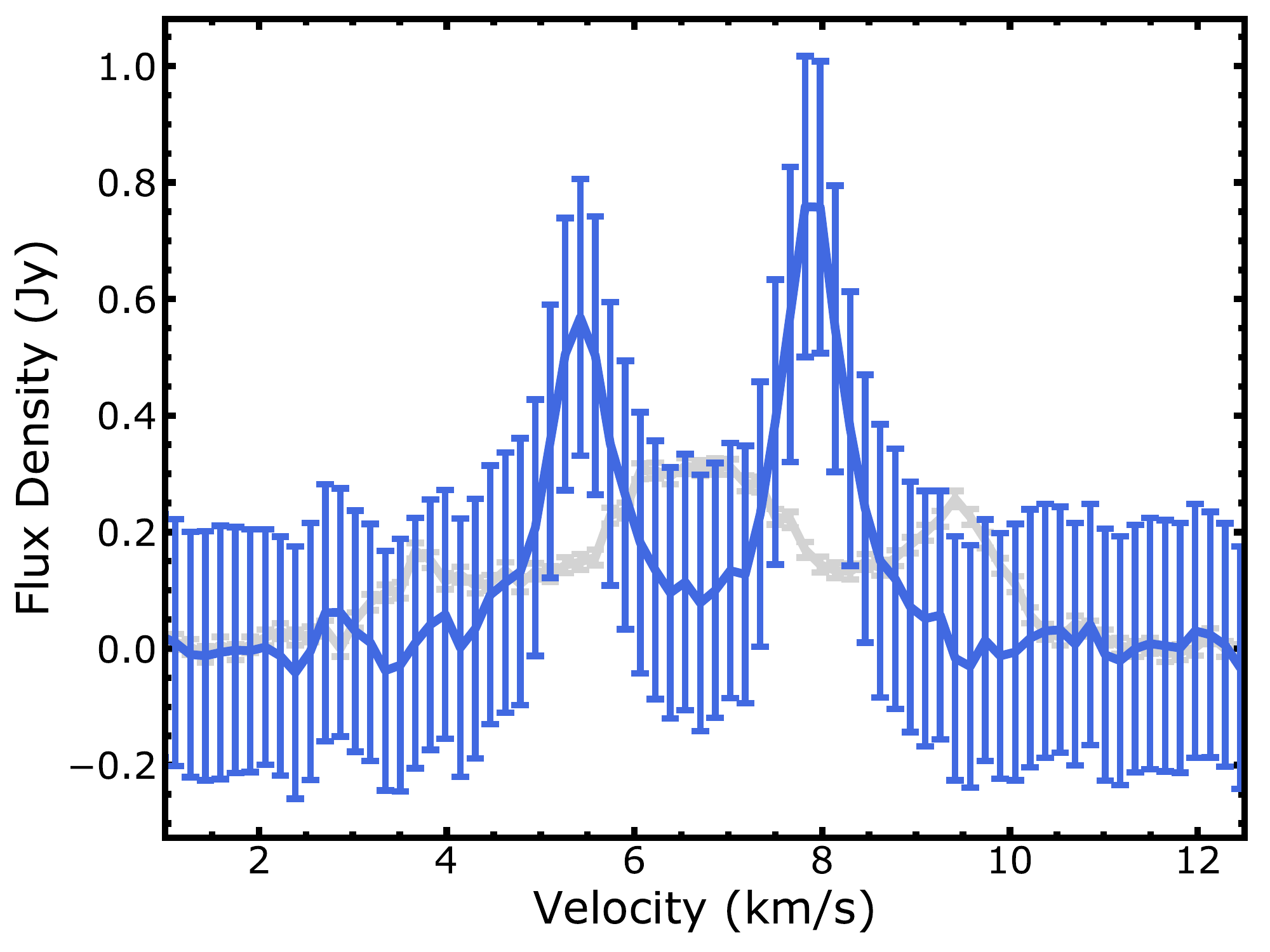}
\end{center}
\caption{Spectra (flux density) of $^{12}$CO $J$=2-1 (left) and C$_2$H $N$=3-2, $J$=$\frac{7}{2}$-$\frac{5}{2}$, where the shifted and stacked spectra are integrated over an elliptical region with the disk geometrical parameters and outer radius of $4\arcsec$ and $3\arcsec$, respectively. The intensity maps are produced with the parameters specified in Table~\ref{tab:fluxes}. The grey lines show the original disk spectra extracted from the same area, with the typical double horn profile apparent in the $^{12}$CO line. Two hyperfine components are clearly visible in the C$_2$H emission. The systemic velocity is 5.5\,km\,s$^{-1}$ \citep{keppler19}.}
\label{fig:spectra}
\end{figure*}

\subsection{Line fluxes}
\label{sec:fluxes}

The spectral setup was built to target simple molecules, since complex organic molecules (COMs) are difficult to detect in in protoplanetary disks due to the low temperatures and large partition functions of the numerous transitions. As described in Sec.~\ref{sec:introduction}, the molecules are divided in five groups: CO isotopologues and hydrogenated versions of CO ($^{12}$CO, $^{13}$CO, C$^{18}$O, H$_2$CO, CH$_3$OH), hydrocarbons (C$_2$H and c-C$_3$H$_2$), cyanides (H$^{13}$CN, HC$^{15}$N, DCN), ions (HCO$^+$, H$^{13}$CO$^+$) and S-bearing molecules (CS, SO). A list of the main molecular lines covered by the spectral setup is reported in Table~\ref{tab:spectral_setup}.

\begin{figure*}
\begin{center}
\begin{tabular}{c}
\includegraphics[width=\textwidth]{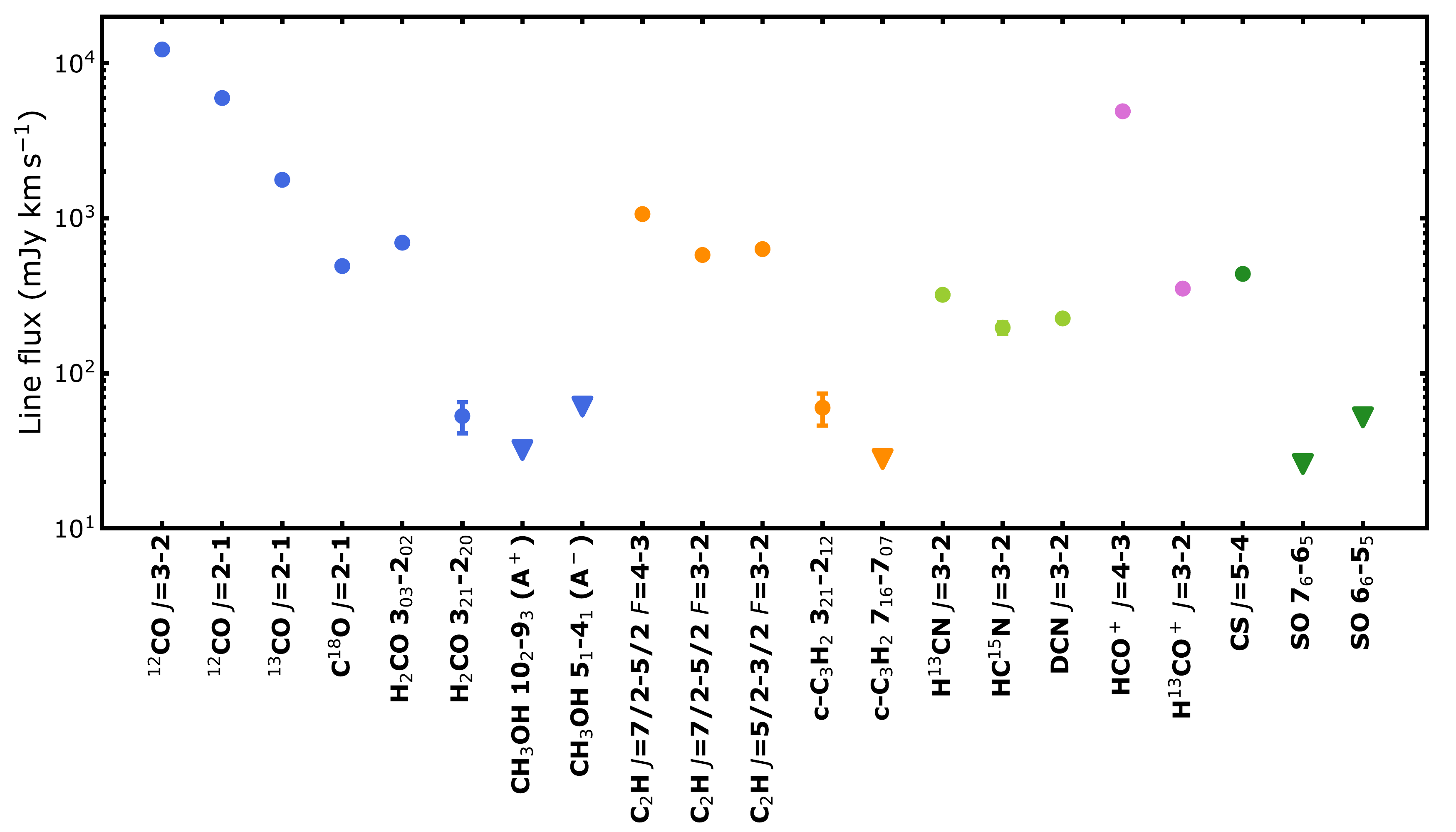}\\
\end{tabular}
\end{center}
\caption{Line fluxes and upper limits, portrayed as triangles. The corresponding values are reported in Table~\ref{tab:fluxes}. Errorbars represent the 1$\sigma$ intervals, while upper limits are shown at 3$\sigma$, where $\sigma$ is computed as the rms of the spectrum times the square root of the number of velocity channels used in the integration. Flux calibration uncertainties are not included. In most cases, the uncertainty is smaller than the marker size. Colors indicate the five different groups of molecules included in this study: CO isotopologues and hydrogenated versions of CO ($^{12}$CO, $^{13}$CO, C$^{18}$O, H$_2$CO, CH$_3$OH), hydrocarbons (C$_2$H and c-C$_3$H$_2$), cyanides (H$^{13}$CN, HC$^{15}$N, DCN), ions (HCO$^+$, H$^{13}$CO$^+$) and S-bearing molecules (CS, SO).}
\label{fig:fluxes}
\end{figure*}

\begin{figure*}
\begin{center}
\begin{tabular}{c}
\includegraphics[width=\textwidth]{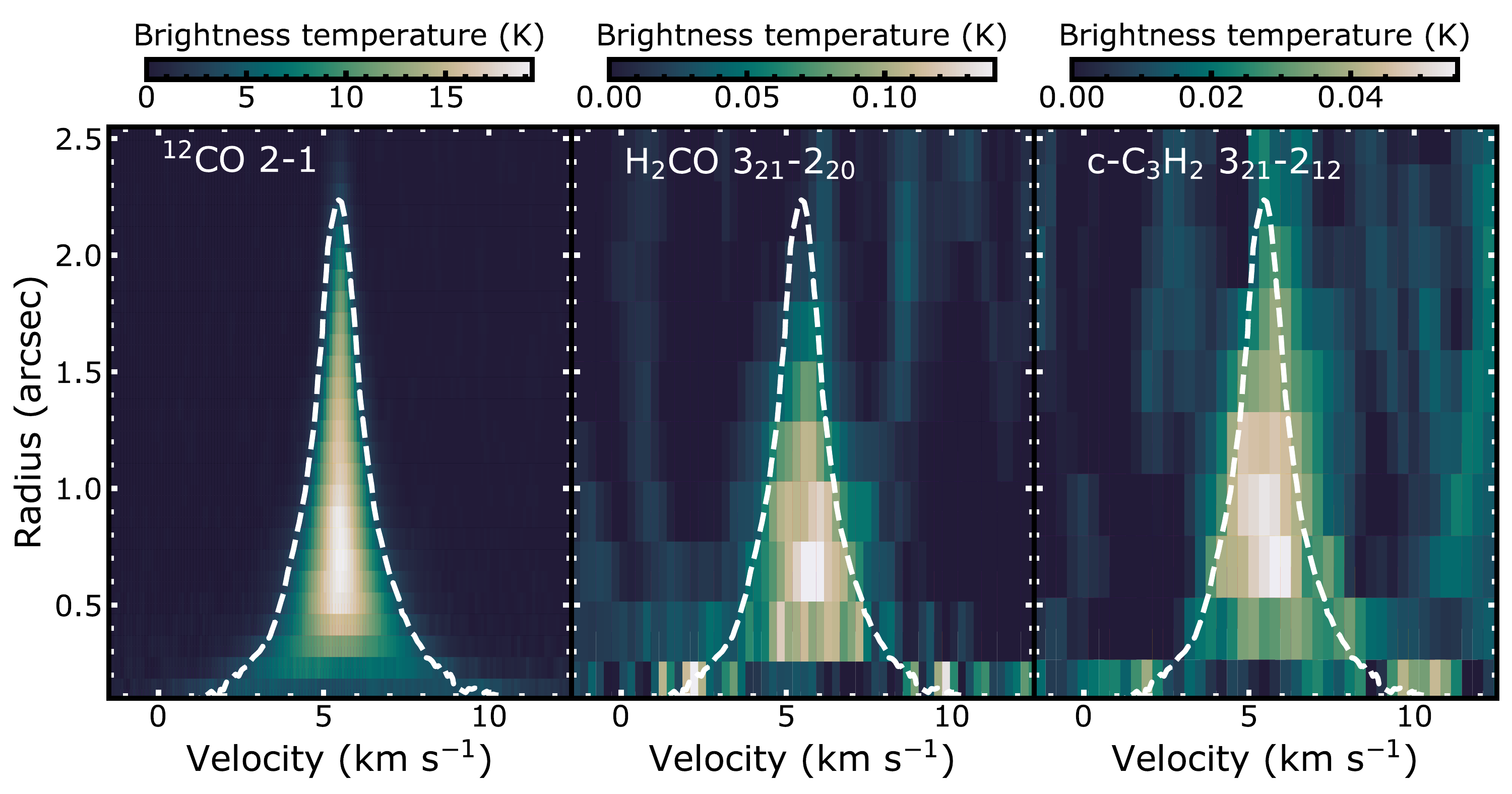}\\
\end{tabular}
\end{center}
\caption{Azimuthally averaged spectra of three molecular lines, showing the detection of H$_2$CO 3$_{21}$-2$_{20}$ and c-C$_3$H$_2$ 3$_{21}$-2$_{12}$, and the $^{12}$CO $J$=2-1 line as comparison. The two weak lines are imaged with natural weighting, and an additional $uv$-tapering to increase sensitivity to extended emission. The 2D spectra are shown in brightness temperature, with the conversion performed under the Rayleigh Jeans approximation. The spectra are computed after de-projecting and shifting the velocities using the kinematical and geometrical model by \citet{keppler19}, and then averaging the spectra along the disk annuli. The H$_2$CO and c-C$_3$H$_2$ lines have been super-sampled by a factor of four in velocity. Dashed white contours show the 3\,K level of the $^{12}$CO emission.}
\label{fig:teardrop}
\end{figure*}

After imaging all lines with weighting and channel widths as reported in Table~\ref{tab:fluxes}, we employ a stacking method to boost the snr of the lines. To do so, we use the same technique used in \citet{Teague16}, similar to the one developed by \citet{yen2016} and \citet{matra2017}. Exploiting the kinematical fit of the Keplerian disk obtained by \citet{keppler19}, at every spatial location we de-project the emission and then shift the local spectrum to the systemic velocity using the \texttt{GoFish} package \citep{gofish}. We then integrate the spectra within $3\arcsec$ from the disk center ($4\arcsec$ for the $^{12}$CO $J$=2-1 line). Fig.~\ref{fig:spectra} shows the integrated spectrum of the $^{12}$CO $J$=2-1 and of the C$_2$H $N$=3-2, $J$=$\frac{7}{2}$-$\frac{5}{2}$ fine structure line as examples, where in the latter the two hyperfine components $F$=4-3 and $F$=3-2 are clearly distinguishable. To obtain line fluxes, we spectrally integrate the spatially integrated stacked spectrum between 3-8\,km\,s$^{-1}$, with the systemic velocity being $5505\pm2\,$m\,s$^{-1}$ \citep{keppler19}. The two $^{12}$CO and the HCO$^+$ lines are integrated between 1.5-9.5\,km\,s$^{-1}$, as high-velocity emission is observed in the channel maps (Section \ref{sec:moment0}). Similarly, the $^{13}$CO, H$^{13}$CN and C$_2$H lines are integrated between 2-9\,km\,s$^{-1}$. The C$_2$H $J$=$\frac{7}{2}$-$\frac{5}{2}$ hyperfine structure lines are partially blended (see Fig.~\ref{fig:spectra}), and fluxes have been estimated by fitting a double Gaussian to the profile. Only the $F$=3-2 hyperfine component of the C$_2$H $J$=$\frac{5}{2}$-$\frac{3}{2}$ is covered by the spectral setup. Since only half of the line is detected at the edge of the spectral window, we consider only the part of the disk within $80^\circ$ of the red-shifted major axis and then assume azimuthal symmetry to extract the total flux. The flux uncertainty for all lines is estimated as standard deviation of the velocity bins not included in the velocity integration, times the square root of the number of bins over which we integrate. The line fluxes and upper limits are reported in Table~\ref{tab:fluxes}, and shown in Figure~\ref{fig:fluxes}.

We detect 14 lines in our Band 6 ($\sim217$-$262$\,GHz) program, and with the two additional $^{12}$CO $J$=3-2 and HCO$^{+}$ lines from the Band 7 ($\sim345$-$357$\,GHz) data, we detect a total of 16 lines, from 12 species (including isotopologues). We detect multiple transitions of $^{12}$CO, H$_2$CO and C$_2$H. The detected lines span a large range of fluxes, with more than two orders of magnitude between the bright $^{12}$CO 3-2 line ($12.3\pm0.1\,$Jy\,km\,s$^{-1}$) and the faint c-C$_3$H$_2$ 3$_{21}$-2$_{12}$ line ($60\pm14$\,mJy\,km\,s$^{-1}$).

\begin{figure*}
\begin{center}
\begin{tabular}{c}
\includegraphics[width=0.9\textwidth]{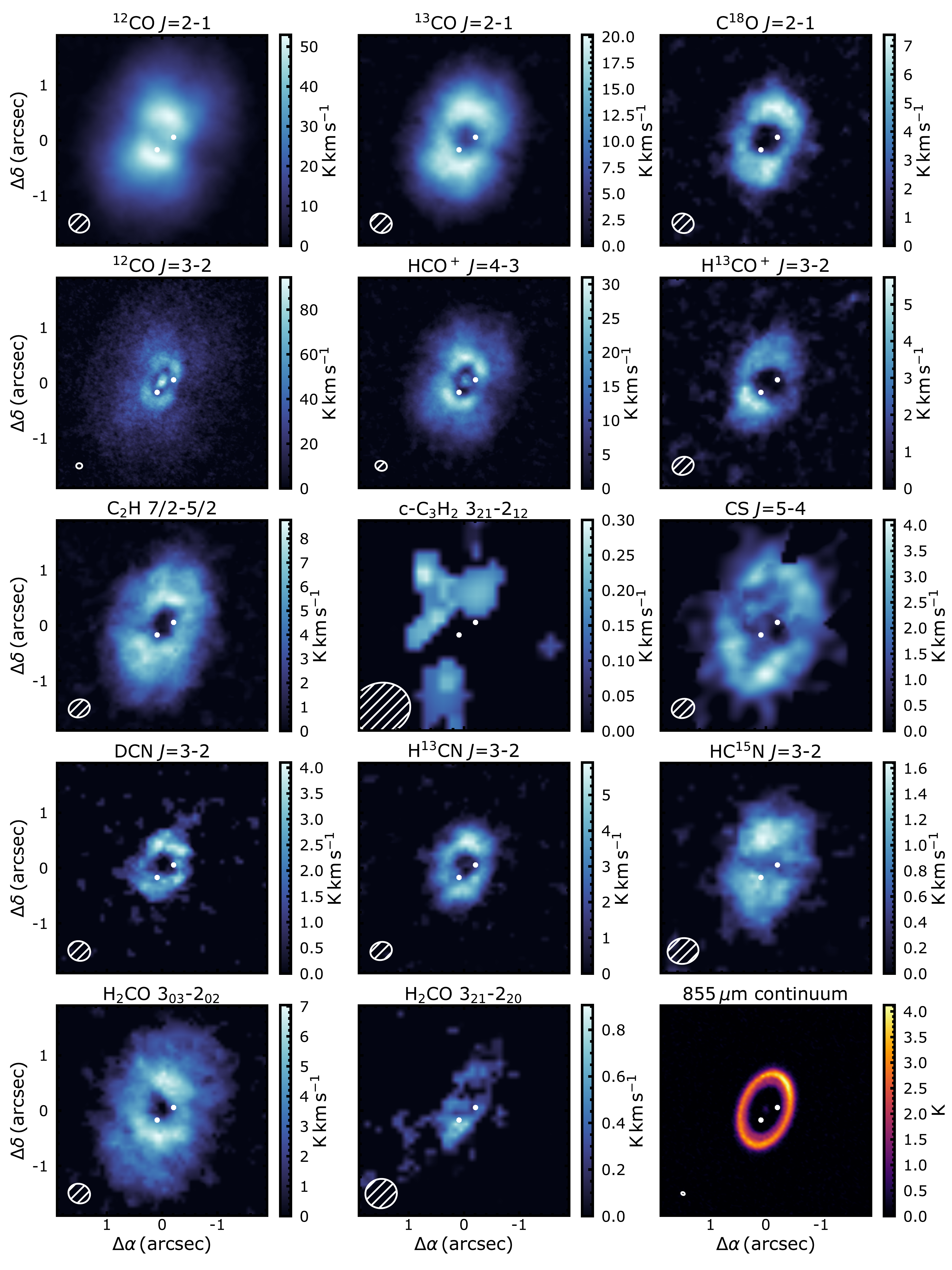}\\
\end{tabular}
\end{center}
\caption{Integrated intensity maps of the 14 detected lines, and of the 885$\,\mu$m continuum from \citet{isella19} in the bottom right in comparison. The imaging parameters of all lines are listed in Table~\ref{tab:images}. The CS and H$_2$CO 3$_{21}$-2$_{20}$ channel maps have been collapsed using a Keplerian mask, while all others have been produced with a $2.5\sigma$ clip.}
\label{fig:moment0}
\end{figure*}

The c-C$_3$H$_2$ 3$_{21}$-2$_{12}$ and  H$_2$CO 3$_{21}$-2$_{20}$ lines are only marginally detected, with a snr of 4.3 and 4.4, respectively. In order to increase the significance of the detections, we image the lines with an additional $uv$-tapering of $0.7\arcsec$. While this reduces the sensitivity to point sources, the down-weight of long baselines can increase the sensitivity to extended structure. For c-C$_3$H$_2$, the resulting beam is $1\farcs06\times0\farcs92$ (PA=$104.6^\circ$) with an rms of $0.9\,$mJy\,beam$^{-1}$ over one 1.38\,km\,s$^{-1}$ channel, while for H$_2$CO, the synthesized beam is $1\farcs03\times0\farcs93$ (PA=$92,0^\circ$), with an rms of 1.2\,mJy\,beam$^{-1}$ over one 1\,km\,s$^{-1}$ channel. The c-C$_3$H$_2$ 3$_{21}$-2$_{12}$ line shows a $5.4\sigma$ detection in the channel map at $6.45\,$km\,s$^{-1}$, with another adjacent channel showing a $4\sigma$ detection. The H$_2$CO 3$_{21}$-2$_{20}$ presents a $8.3\sigma$ detection at 8\,km\,s$^{-1}$, with four adjacent channels showing signal at $\gtrsim4\sigma$. Both channel maps are included in the Appendix (Fig.~\ref{fig:channel_maps}). As a final test to assess the robustness of the detection of the two lines, we divide the projected emission in concentric annuli, separated by one quarter of the major axis of the beam. Along these annuli, we compute the azimuthally averaged spectrum after shifting the velocities to the systemic one by exploiting the kinematical information. This is performed using the \texttt{GoFish} package \citep{gofish}. The azimuthally averaged spectra are shown in Fig.~\ref{fig:teardrop}, together with the azimuthally averaged spectrum of $^{12}$CO $J$=2-1 for comparison. Both lines clearly show an emission peak at the systemic velocity, indicating again that both lines are robustly detected. It is worth noting that thanks to this method we manage to obtain a detection at the 50\,mK level in brightness temperature units.

We do not detect SO or methanol. The three lines of sulphur monoxide are SO 6$_6$-5$_5$, 7$_6$-6$_5$ and 2$_3$-3$_2$, with upper energies of 56.5, 47.5 and 21.0\,K (see Table~\ref{tab:column_densities}), with the last being fainter by two orders of magnitude. None of the lines are detected in the channel maps or in the azimuthally averaged spectra of the natural images with and without $uv-$tapering. One line of $^{34}$SO is also included in the spectral setup at 246.6636\,GHz, but is not detected, as expected from the non-detection of the main isotopologue. The two brightest methanol lines included in the spectral setup are both for A-type states (analogue to ortho-ammonia in the spin state of the methyl group).  The transitions are $10_2$ - $9_3$(A$^+$) at 232.4185 GHz and $5_1$ - $4_1$(A$^{-}$) at 243.9157 GHz. Neither of the two lines are detected with the methods described above. The last notable non detection is for the c-C$_3$H$_2$ 7$_{16}$-7$_{07}$ at 218.7327\,GHz. Finally, a matched filter analysis is performed on the whole spectral setup in the $uv$-plane, with particular attention given to the wide basebands. To do so, we exploit the \texttt{VISIBLE} package by \citet{VISIBLE}, which has been successfully applied to protoplanetary disks observations \citep[e.g.,][]{loomis2018,loomis2020}. This method requires a model image as matched filter, and we thus run the analysis onto the continuum subtracted visibilities with the CS and H$^{13}$CO$^+$ channel maps as filters. No new additional line is detected with this method.

\subsection{Integrated maps}
\label{sec:moment0}

Integrated intensity maps of the 14 detected transitions are derived with the same imaging parameters used in Section~\ref{sec:fluxes}. The c-C$_3$H$_2$ 3$_{21}$-2$_{12}$ line is imaged with natural weighting and an additional $uv$-tapering, as specified in Section~\ref{sec:fluxes}, to increase brightness temperature sensitivity. Integrated intensity maps of the 14 detected lines are then created integrating over the whole cubes with a $2.5\sigma$ clipping. In order to increase the contrast of the images for faint lines, we adopt a Keplerian masking technique, by masking any signal that does not originate from the Keplerian mask used in the cleaning (see Section~\ref{sec:reduction}). This technique has already been exploited by different groups to boost the snr in their data \citep[e.g.,][]{salinas2017,ansdell2018,bergner2018}. The cases where the image quality improves significantly are CS and H$_2$CO 3$_{21}$-2$_{20}$. The emission morphology does not vary significantly between the different techniques for the other lines.

The integrated intensity maps are shown in Fig.~\ref{fig:moment0}, where the molecular emission is compared to the high resolution 855$\,\mu$m continuum emission from \citet{isella19} (cleaned with Briggs weighting, robust=0.3). All lines show a bright ring of emission outside the orbit of the two massive planets, which are highlighted as white dots in the figure. While the general spatial morphology is similar among different lines, they clearly exhibit different structures, together with very different intensities. For example, hydrogen cyanide isotopologues and H$^{13}$CO$^+$ show a very defined narrow ring similarly to the continuum emission, while H$_2$CO and C$_2$H exhibit more extended emission. We analyze the radial profiles of the emission lines in Section~\ref{sec:radial_profiles}. Most lines seem to show very azimuthally symmetric emission, with the only possible exception of H$^{13}$CO$^+$. 

The high resolution images of the Band 7 data of $^{12}$CO $J$=3-2 and HCO$^+$ $J$=4-3 show an inner disk, inside the deep gap carved by the planets \citep[][]{Long18,keppler19}. Similarly to the Band 6 continuum emission, it is difficult to distinguish any inner disk in the other molecular tracers, due to the lower angular resolution. However, we can explore the contribution of molecular gas within the orbits of the planets in the high velocity channels, associated to emission from small radii. From \citet{wang_ea_2021}, planet c has an orbit with semi-major axis of 33.2\,au. Using the geometrical parameters by \citet{keppler19}, in particular the disk inclination of $51.7^\circ$ and a stellar mass of $0.87\,M_\odot$, any Keplerian gas with projected velocity higher than 3.75\,km\,s$^{-1}$ from the systemic velocity must emit from within the orbit of PDS~70c. With a systemic velocity of 5.5\,km\,s$^{-1}$, this includes emission at $<1.75\,$km\,s$^{-1}$ and $>9.25\,$km\,s$^{-1}$. The $^{12}$CO $J$=2-1 line shows emission at high velocities (at $\sim0$ and $\sim10\,$km\,s$^{-1}$, see Figure ~\ref{fig:spectra}), indicating a contribution from small radii, as is the case for the $^{12}$CO $J$=3-2 and the HCO$^+$ $J$=4-3 lines. Emission lines of C$_2$H, $^{13}$CO and H$^{13}$CN start to show detectable emission at $\sim1.85\,$km\,s$^{-1}$, thus indicating that no emission is likely seen within the orbit of the outermost detected planet at the sensitivity of our observations.

\begin{deluxetable*}{lcccccc} 
	\tablecaption{Images used to extract the radial profiles of the 14 detected lines. The radius of peak indicates the radius at which the radial profile of the integrated intensities has a maximum at $r>0.2\arcsec$. \label{tab:images}}
	\tablecolumns{7} 
	\tablewidth{\textwidth} 
	\tablehead{
		\colhead{Species}                          &
		\colhead{Transition}                       &
		\colhead{uv-tapering}                     & 
        \colhead{Beam (PA)}                  & 
        \colhead{Radius of peak}          &
        \colhead{Error$^a$}          &
        \colhead{Peak integrated intensity$^b$} \\
		\colhead{}        & 
		\colhead{}        &
		\colhead{}        &
		\colhead{}          &
		\colhead{($\arcsec$)}        &
		\colhead{($\arcsec$)}        &
		\colhead{(K\,km\,s$^{-1}$)}      
		  }
\startdata
$^{12}$CO & 3-2 & n & 0$\farcs$11$\times$0$\farcs$10  (90.3$^\circ$)                    & 0.40  &  0.03  &  57.1 \\
$^{12}$CO  & 2-1 & n & 0$\farcs$37$\times$0$\farcs$33 (67.1$^\circ$)             & 0.42  &  0.09  &  46.0 \\
$^{13}$CO  & 2-1 & n & 0$\farcs$39$\times$0$\farcs$35 (69.6$^\circ$)             & 0.54  &  0.10  &  17.0 \\
C$^{18}$O  & 2-1 & n & 0$\farcs$39$\times$0$\farcs$35 (68.3$^\circ$)             & 0.64  &  0.10  &  5.6 \\
H$_2$CO  & 3$_{03}$-2$_{02}$ & n & 0$\farcs$40$\times$0$\farcs$35 (69.0$^\circ$) & 0.55  &  0.10  &  5.4 \\
H$_2$CO  & 3$_{21}$-2$_{20}$ & n & 0$\farcs$57$\times$0$\farcs$52 (100.8$^\circ$)     & 0.21  &  0.14  &  0.4 \\
C$_2$H$^c$  & 3-2 $J$=$\frac{7}{2}$-$\frac{5}{2}$ & n & 0$\farcs$39"$\times$0$\farcs$32 (107.7$^\circ$) & 0.63  &  0.10  &  6.8 \\
c-C$_3$H$_2$  & 3$_{21}$-2$_{12}$ & y & $1\farcs06\times0\farcs92$ (104.6$^\circ$) & 0.93  &  0.27  &  0.1 \\
H$^{13}$CN  & 3-2 & n & 0$\farcs$39$\times$0$\farcs$32 (107.1$^\circ$)           & 0.44  &  0.10  &  4.4 \\
HC$^{15}$N  & 3-2 & n & 0$\farcs$56$\times$0$\farcs$46 (100.6$^\circ$)                & 0.49  &  0.14  &  1.1 \\
DCN  & 3-2 & n & 0$\farcs$40$\times$0$\farcs$35 (70.8$^\circ$)                   & 0.45  &  0.10  &  2.7 \\
HCO$^+$  & 4-3 & n & 0$\farcs$21$\times$0$\farcs$18 (73.1$^\circ$)                    & 0.45  &  0.05  &  23.4 \\
H$^{13}$CO$^+$  & 3-2 & n & 0$\farcs$39$\times$0$\farcs$32 (108.8$^\circ$)       & 0.54  &  0.10  &  3.3 \\
CS  & 5-4 & n & 0$\farcs$42$\times$0$\farcs$34 (108.8$^\circ$)                   & 1.09  &  0.10  &  2.7 \\
\enddata
\tablenotetext{}{$^a$ Error is estimated as one quarter of the major axis of the clean beam; $^b$ Integrated intensity at the peak radius for $r>0\farcs2$ in K\,km\,s$^{-1}$, converted from intensity units assuming Rayleigh-Jeans regime; $^c$ the two hyperfine components have been considered simultaneously.}
\end{deluxetable*}

\subsection{Radial profiles}
\label{sec:radial_profiles}

\begin{figure*}
\begin{center}
\begin{tabular}{c}
\includegraphics[width=0.95\textwidth]{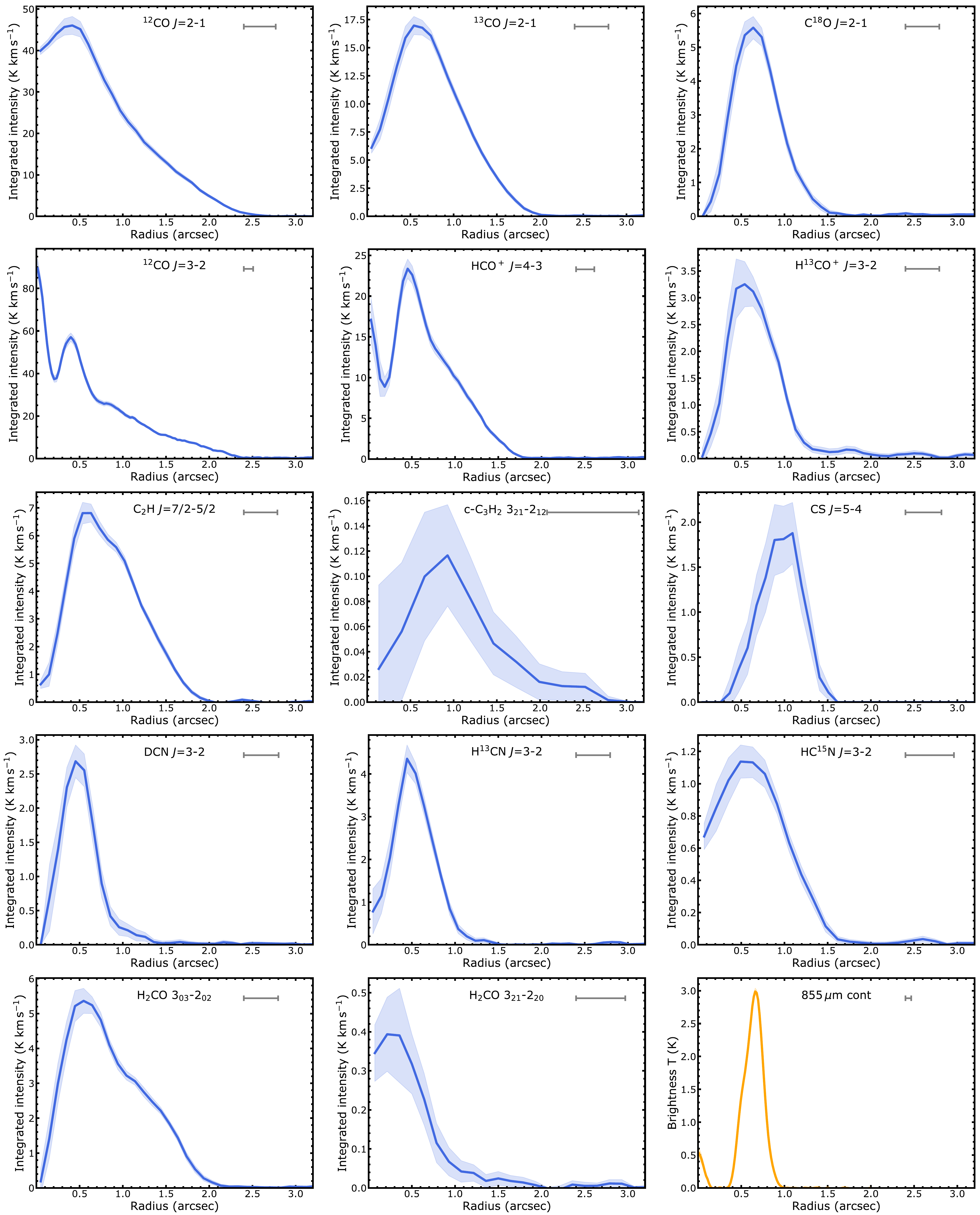}\\
\end{tabular}
\end{center}
\caption{Azimuthally averaged radial profiles of the 14 detected lines, and of the 885$\,\mu$m continuum by \citet{isella19} in the bottom right in comparison. The imaging parameters of the respective channel maps are listed in Table~\ref{tab:images}, with integrated maps images shown in Figure~\ref{fig:moment0}. The ribbon shows the $1\sigma$ rms, excluding the uncertainty from absolute flux calibration. The grey line at the top right of all panels shows the beam major axis. The brightness temperature conversion was done under the Rayleigh-Jeans approximation.}
\label{fig:radial_profiles}
\end{figure*}

\begin{figure*}
\begin{center}
\begin{tabular}{c}
\includegraphics[width=0.45\textwidth]{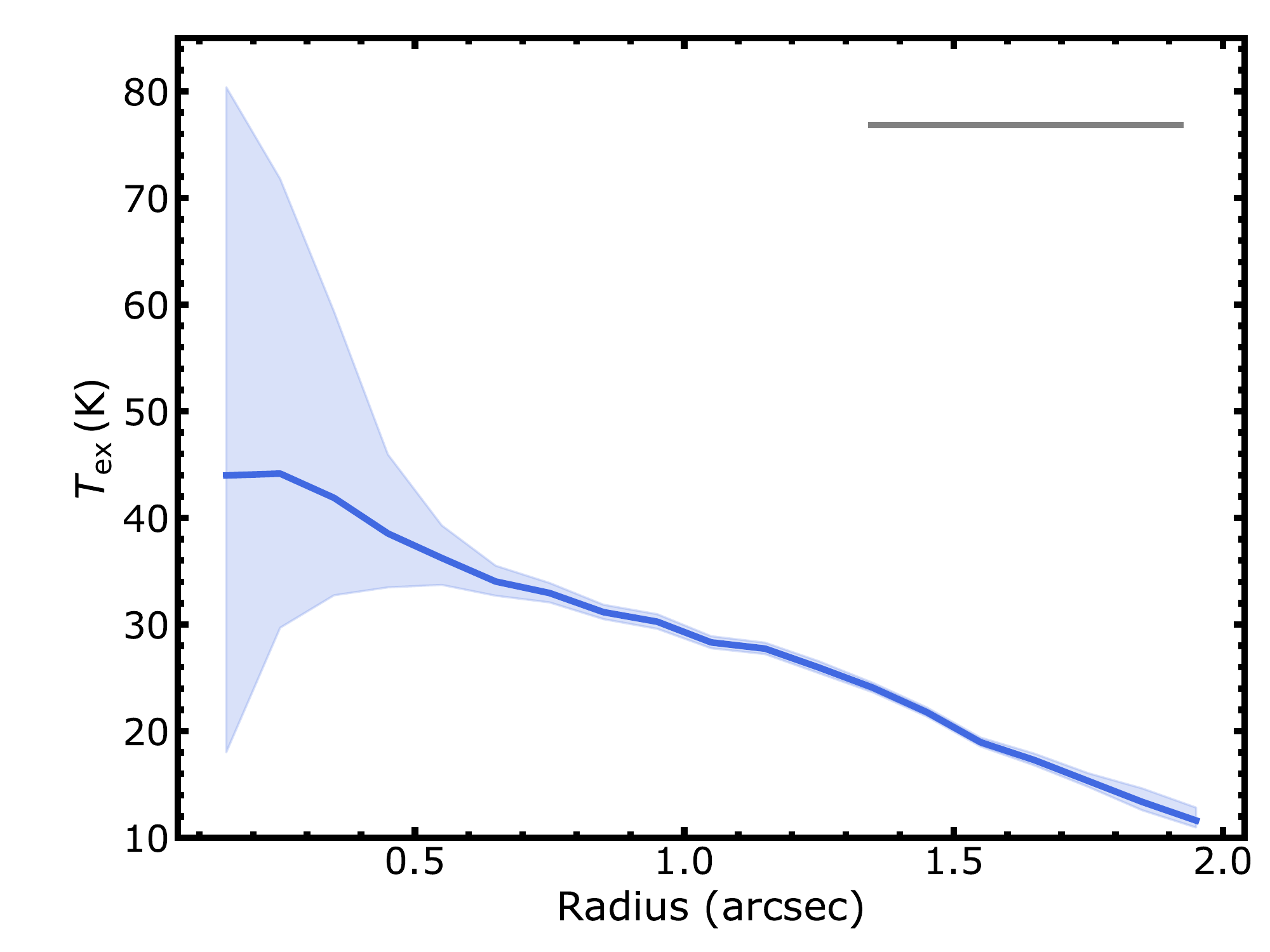}
\includegraphics[width=0.45\textwidth]{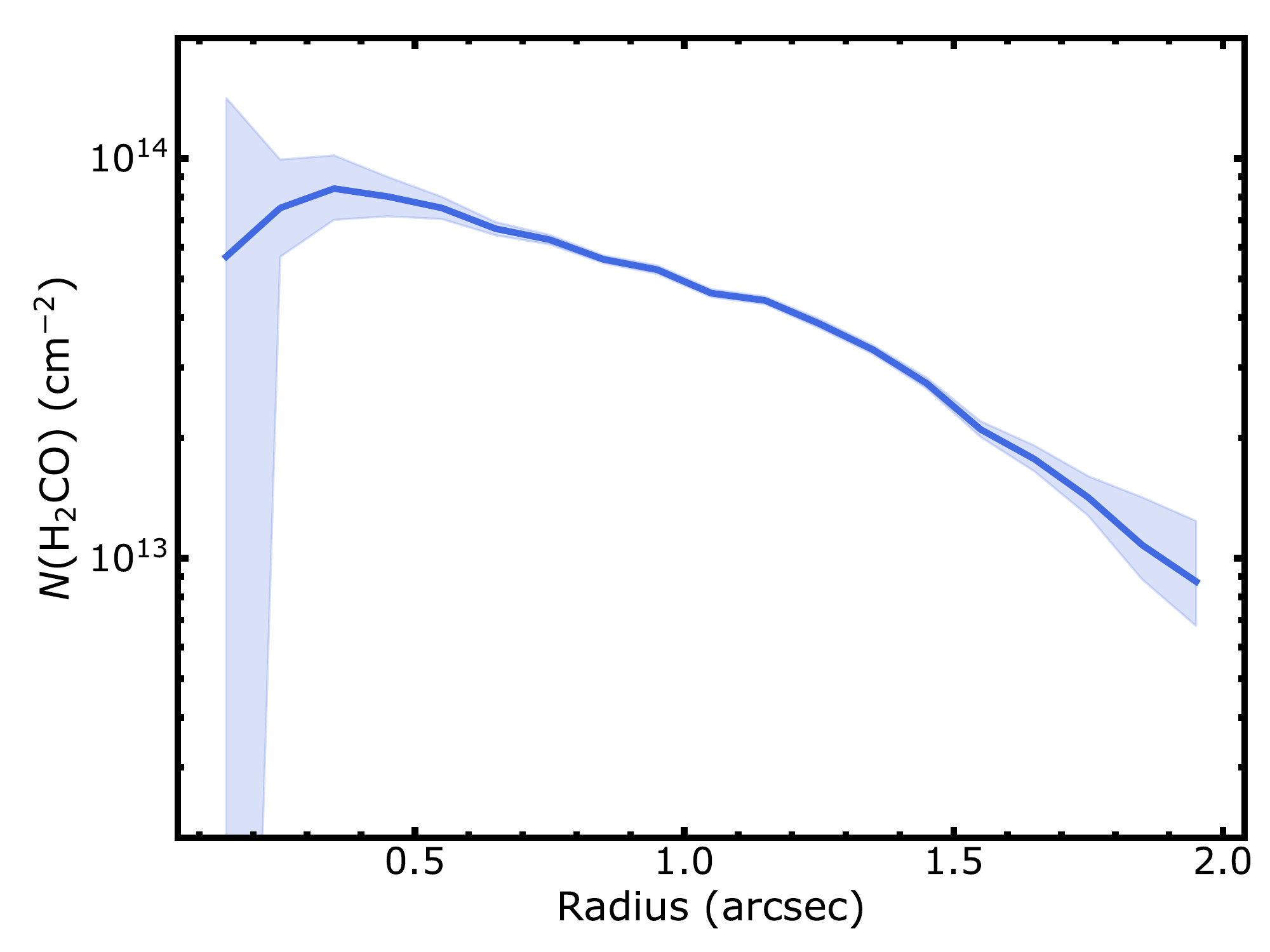}\\
\end{tabular}
\end{center}
\caption{Left: H$_2$CO excitation temperature derived from rotational diagram analysis of the 3$_{03}$-2$_{02}$ and 3$_{21}$-2$_{20}$ transitions. The grey line at the top right shows the beam major axis. Right: H$_2$CO column density. Integrated intensity maps with natural weighting were used in the analysis to allow for enough snr in the weaker 3$_{21}$-2$_{20}$ transition.}
\label{fig:h2co}
\end{figure*}

In order to extract the radial profiles of the integrated intensities, we azimuthally average the de-projected integrated intensity maps shown in Fig.~\ref{fig:moment0}. Error bars are estimated from the standard deviations of the emission within the annuli, divided by the square root of the number of independent beams across the azimuthal range. The width of the annuli are a quarter of the beam major axis. The resulting integrated intensity profiles are shown in Figure~\ref{fig:radial_profiles}. 

A deep gap co-located with the orbital radius of PDS 70b is observed in HCO$^+$ and $^{12}$CO $J$=3-2 \citep{Long18,keppler19}. For all other lines, no emission is seen from an inner disk in the integrated intensity profiles, in line with the discussion of Section~\ref{sec:moment0}. The brightest lines clearly exhibit emission at radii $>2\arcsec$, in particular H$_2$CO, $^{12}$CO, $^{13}$CO and C$_2$H. The $^{12}$CO $J$=2-1 line, which has a higher sensitivity than the $J$=3-2 transition, shows emission out to radii $>3\arcsec$. The H$_2$CO line shows a shoulder of emission outside the continuum ring between 1-$1.5\arcsec$, which is also hinted in the DCN line. This may be due to enhanced gas temperature outside the dust ring where photons penetration depths are higher due to the lower dust column densities \citep[e.g.,][]{oberg2015,cleeves2016,facchini2017,rosetta_I}. This would induce a stronger thermal and photo-desorption of CO, leading to the H$_2$CO shoulder.

\section{Derivation of disk physical parameters}
\label{sec:physics}

In the following section, disk fluxes and integrated intensity radial profiles are exploited to derive physical and chemical parameters of the PDS~70 disk, mainly column densities and deuteration and fractionation profiles, without requiring any chemical modeling.

\subsection{Column Densities and Optical Depths}
\label{sec:column}

In the optically thin limit, the column density of an emitting molecule, $N$, can be related to the integrated flux density, $S_{\!\nu} \Delta v$, and excitation temperature, $T_{\rm ex}$, via,

\begin{equation}
\label{eq:column}
    N_{\rm thin} = \frac{4\pi}{A_{\rm ul} h c} \frac{S_{\!\nu} \Delta v}{\Omega} \frac{Q(T_{\rm ex})}{g_{\rm u}} \, \exp\left(\frac{E_{\rm u}}{T_{\rm ex}}\right),
\end{equation}

\noindent where $\Omega$ is the area over which the flux was integrated, $A_{\rm ul}$ is the Einstein coefficient of the transition, $E_{\rm u}$ is the upper state energy, $h$ is the Planck constant, $c$ the speed of light in vacuum, $g_{\rm u}$ the upper state degeneracy and $Q$ the partition function. When the optical depth, $\tau$, is no longer negligible, this can be corrected by, 

\begin{equation}
    N = \frac{N_{\rm thin} \, \tau}{1 - \exp(-\tau)} = N_{\rm thin} C_{\tau}
\end{equation}

\noindent following, for example, \citet{GoldsmithLanger1999}. In turn, the optical depth at the line center (assuming a top hat line profile) can be calculated via, 

\begin{equation}
    \tau = \frac{A_{\rm ul} c^3 }{8 \pi \nu^3 \Delta v} Ng_u \frac{\exp(-E_{\rm u} / T_{\rm ex})}{Q(T_{\rm ex})} \left[ \exp\left( \frac{h \nu}{k T_{\rm ex}} \right) - 1 \right]
\end{equation}

\noindent where $\nu$ is the frequency of the transition. In the following, we use the molecular data provided in Table~\ref{tab:column_densities}, taken from the CDMS catalogue \citep{cdms}, and interpolate the partition function, $Q$, with a cubic spline.

\subsubsection{H$_2$CO}

As we detected two transitions of H$_2$CO, $3_{03}$-$2_{02}$ and $3_{21}$-$2_{20}$, we are able to constrain both $N$ and $T_{\rm ex}$ simultaneously \citep[e.g.,][for similar analyses]{salinas2017,bergner2019,rosetta_I}. For the integrated fluxes, we radially bin the integrated intensity profiles shown in Fig.~\ref{fig:radial_profiles} into $0\farcs1$ wide annuli to measure $S_{\!\nu} \Delta v / \Omega$.

Following, e.g., \citet{loomis2018}, we use \texttt{emcee} \citep{Foreman-Mackey+13} to explore the posterior distributions for $\{N({\rm H_2CO}),\, T_{\rm ex}\}$ while self-consistently solving for the optical depth of the two transitions. For each radial bin, we use 128 walkers, each taking 1,000 steps to burn in, and then an additional 500 steps to sample the posterior distributions. As the posteriors were found to be Gaussian, we take the median value for each parameter as the most likely value, and the 16th to 84th percentile as a quantification of the uncertainty on this. The resulting radial profiles of $N({\rm H_2CO})$ and $T_{\rm ex}$ are found in Fig.~\ref{fig:h2co}.

The excitation temperature ranges between $\sim10-40\,$K, with the temperature decreasing in the outer disk regions. Higher temperatures are found for the inner regions, as expected for the cavity wall being strongly irradiated, however the snr of the data is low, as reflected by the large inferred uncertainties. Column densities across the $0.5-2\arcsec$ interval range between $10^{13}-10^{14}\,$cm$^{-2}$, in line with other protoplanetary disks \citep[e.g.,][]{pegues2020,rosetta_II}.

\subsubsection{C$_2$H}

Three hyperfine transitions of C$_2$H were detected, however the two 3-2 $J = \frac{7}{2} - \frac{5}{2}$ transitions were sufficiently blended that typical hyperfine-fitting methods, e.g., \citet{hily-blant2013b,hily-blant_ea_2017,teague_loomis_2020}, could not be used. Similarly, there was only a single 3-2 $J = \frac{5}{2} - \frac{3}{2}$ fine structure group component detected, precluding typical hyperfine analyses. Instead we follow a similar approach to that used for H$_2$CO, but for each individual component separately, which implicitly assumes that local thermodynamic equilibrium (LTE) is valid.

As the $J = \frac{7}{2} - \frac{5}{2}$ transitions were blended, we used \texttt{GoFish} to calculate azimuthally-averaged, velocity-corrected spectra for annuli spaced $0\farcs1$ apart (as in the right panel of Fig.~\ref{fig:fluxes}). To each of these spectra, we fit two Gaussian components using the \texttt{scipy.optimize.curve\_fit} routine, fixing the line centers to the rest frequency of the two transitions, and assuming that the lines are optically thin such that the combined spectrum is a simple sum of Gaussian components. For the single $J = \frac{5}{2} - \frac{3}{2}$ transition, only part of the disk emission was covered (see discussion in Section~\ref{sec:fluxes}). Thus the azimuthally-averaged, velocity-corrected spectra only considered emission within $80\degr$ of the redshifted major-axis. To this spectrum, a single Gaussian component was fit in order to measure the line flux.

\begin{figure}
\includegraphics[width=\columnwidth]{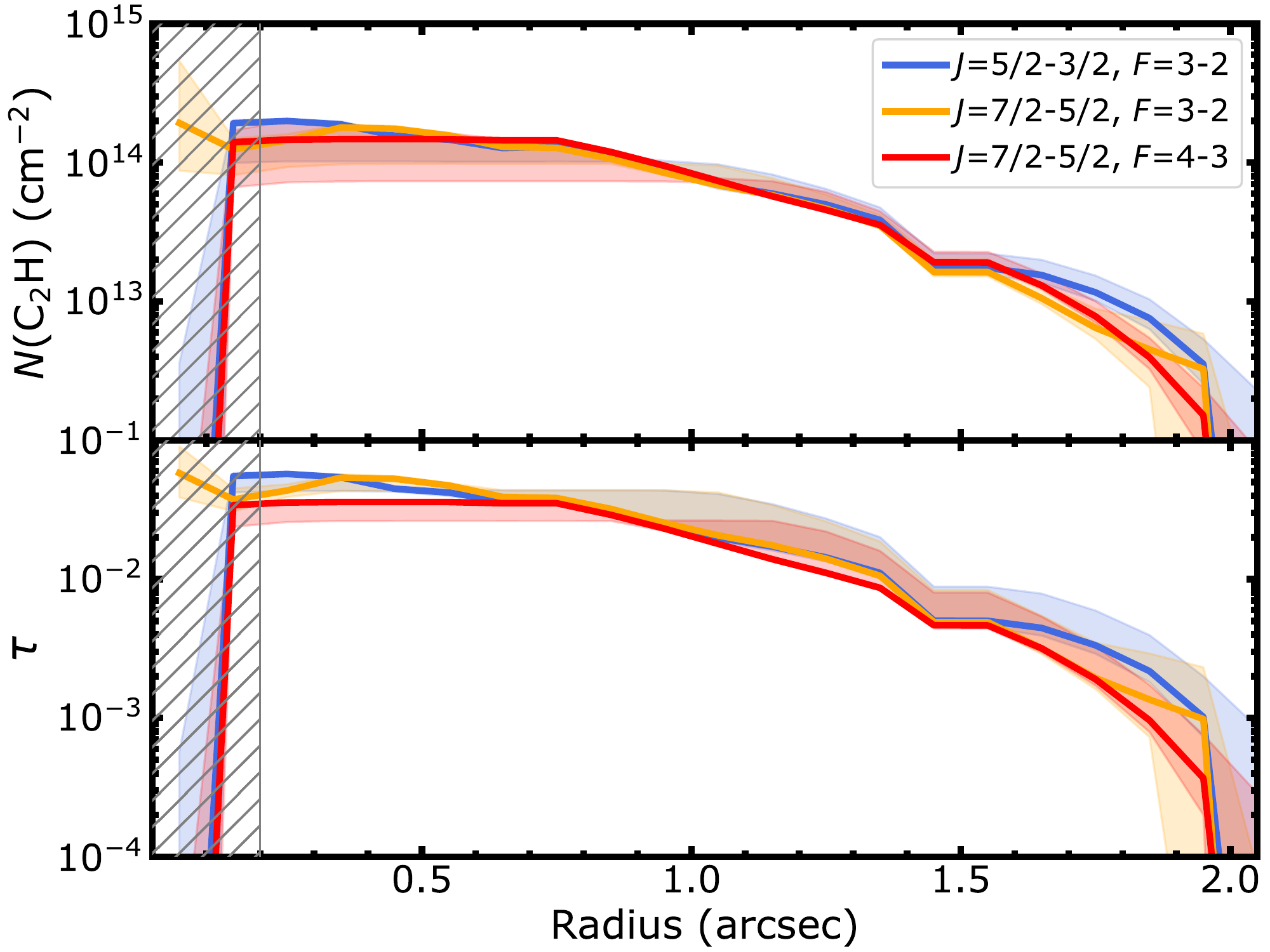}\\
\caption{Derived column densities and optical depths for the three hyperfine C$_2$H lines included in the spectral setup. The assumed excitation temperature $T_{\rm ex}$ is varied between 20-50\,K, with the solid lines showing the 30\,K case. The ribbons include the scatter in the obtained column densities from the posterior distributions between the 16 and 84 percentiles after accounting for the 10\% flux calibration uncertainty, and the systematic uncertainty driven by the unknown $T_{\rm ex}$. The hatched area at $r<0\farcs2$ indicates the region where beam smearing dominates the column density reconstruction.}
\label{fig:c2h}
\end{figure}

As only a single transition is considered, $T_{\rm ex}$ was held constant at values of $\{20,\, 30,\, 50\}$~K, guided by previous analyses of C$_2$H in other sources \citep[e.g.,][]{bergner2019} in order to explore likely ranges of $N({\rm C_2H})$. Again, \texttt{emcee} was used to sample the posterior distribution of $N({\rm C_2H})$, with $\tau$ being calculated self-consistently. The 128 walkers took 1,000 steps to burn in and an additional 500 to sample the posteriors. The resulting radial profiles of $N({\rm C_2H})$ are shown for the three hyperfine components in Fig.~\ref{fig:c2h}, where the solid lines show the result for $T_{\rm ex} = 30$~K, and the shaded regions show the ranges when considering lower and higher $T_{\rm ex}$ values. The concordance between these three transitions suggests that the C$_2$H column density is well constrained and that LTE is a reasonable assumption.

\subsubsection{Other Molecules}

\begin{figure*}
\begin{center}
\begin{tabular}{c}
\includegraphics[width=\textwidth]{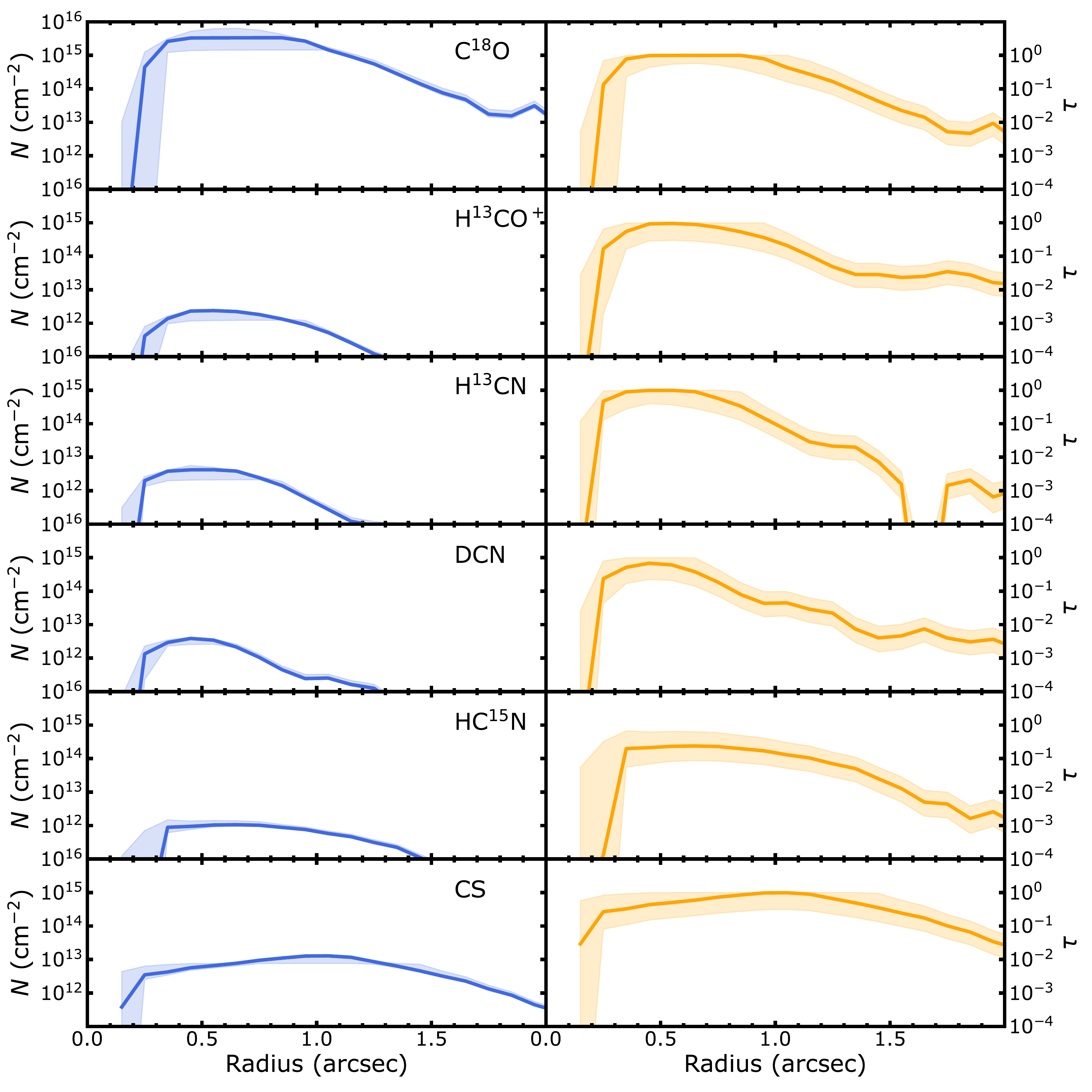}\\
\end{tabular}
\end{center}
\caption{Column densities and optical depths of all detected lines, excluding C$_2$H and H$_2$CO which are reported in Figs.~\ref{fig:h2co}-\ref{fig:c2h}, and c-C$_3$H$_2$ which presents a very weak emission line. The method used to derive both column densities and optical depths is described in Sec.~\ref{sec:column}. The ribbon indicates statistical uncertainties including a 10\% flux calibration uncertainty and systematic uncertainties driven by the unknown excitation temperatures, which have been varied between 20-50\,K. The solid lines represent the profiles for $T_{\rm ex}=30\,$K.}
\label{fig:column_density_profiles}
\end{figure*}

We follow a similar procedure as used for C$_2$H for all other molecules which only have a single transition detected. The integrated intensities are calculated from the azimuthally averaged profiles of the total intensity maps shown in Fig.~\ref{fig:moment0}, and binned into annuli of $0\farcs1$. As with C$_2$H, three $T_{\rm ex}$ values were considered: 20, 30 and 50~K. We add a prior on the optical depth to be $<1$ for all molecules, since all the considered transitions show a brightness temperature that is well below the assumed excitation temperature range. The inferred column density profiles and optical depths are shown in Fig.~\ref{fig:column_density_profiles}.

\subsubsection{c-C$_3$H$_2$ and upper limits on SO and CH$_3$OH}
\label{sec:column_upp_lim}

The detected c-C$_3$H$_2$ line exhibits a low snr, and we thus compute a disk averaged column density by using Eq.~\ref{eq:column} and assuming optically thin emission. A radius of $3\arcsec$ is used to compute $\Omega$. A column density of $5.7^{+4.0}_{-1.6}\times10^{12}$\,cm$^{-2}$ is obtained, including the uncertainty from the unknown excitation temperature (ranged again between 20-50\,K). The relatively high column, associated to the weak molecular emission is easily explained by the large partition function.

From the upper limits on the SO and CH$_3$OH we can derive upper limits on the disk averaged column density following the same approach. The SO 7$_6$-6$_5$ line is the most stringent of the two considered in this work, with a 3$\sigma$ upper limit on the SO disk averaged column density of $1.5\times10^{11}$\,cm$^{-2}$, where the uncertainty of the excitation temperature has been considered as above. As for CH$_3$OH, the 5$_1$-4$_1$(A$^-$) is the most constraining transition, with a relative upper limit on the column density of $6.5\times10^{12}$\,cm$^{-2}$. The 10$_2$-9$_3$(A$^+$) transition leads to much weaker constraint, due to the high upper level energy and lower Einstein coefficient.

\begin{deluxetable*}{lccccccccc} 
	\tablecaption{Molecular line data and peak column densities and optical depths derived in Sec.~\ref{sec:column}. \label{tab:column_densities}}
	\tablecolumns{10} 
	\tablewidth{0.8\textwidth} 
	\tablehead{
		\colhead{Species}                          &
		\colhead{Transition}                       &
		\colhead{$E_{\rm u}$}                     & 
        \colhead{$\log{A_{\rm ul}}$}                  & 
        \colhead{$g_{\rm u}$}           &
        \colhead{$Q(18.75\,$K)}          &
        \colhead{$Q(37.50\,$K)}          &
        \colhead{$Q(75\,$K)}    &
        \colhead{$N_{\rm max}$ $^a$}   &
        \colhead{$\tau_{\rm max}$ $^a$}\\
		\colhead{}        & 
		\colhead{}        &
		\colhead{(K)}        &
		\colhead{(s$^{-1}$)}          &
		\colhead{}        &
		\colhead{}        &
		\colhead{}        &
		\colhead{}        &
		\colhead{(cm$^{-2}$)}       &
		\colhead{}
		  }
\startdata
C$^{18}$O       & 2-1               & 15.8      & -6.2  & 5     & 7.46      & 14.57     & 28.81     & $6.2^{+0.2}_{-2.8} \times10^{15}$ & $9.8^{+0.02}_{-0.41} \times10^{-1}$  \\
H$_2$CO         & 3$_{03}$-2$_{02}$ & 21.0      & -3.55 & 7     & 44.68     & 128.65    & 361.72    & $8.4^{+1.8}_{-1.4} \times10^{13}$ & $8.8^{+11.3}_{-5.8} \times10^{-1}$ \\
H$_2$CO         & 3$_{21}$-2$_{20}$ & 68.1      & -3.80 & 7     & 44.68     & 128.65    & 361.72    & $8.4^{+1.8}_{-1.4} \times10^{13}$ & $1.7^{+0.3}_{-0.3} \times10^{-1}$  \\
CH$_3$OH         & $5_1$ - $4_1$(A$^{-}$) & 49.7      & -3.85 & 11     & 68.75     & 230.24    & 731.07    & $<6.5 \times10^{12}$ & ...$^c$ \\
CH$_3$OH         & $10_2$ - $9_3$(A$^+$) & 165.3      & -4.22 & 21     & 68.75     & 230.24    & 731.07    & $<1.2 \times10^{15}$ & ...$^c$ \\
C$_2$H          & $J$=7/2-5/2 $F$=4-3   & 25.1      & -4.12 & 9    & 37.14     & 72.91     & 144.71    & $2.0^{+0.5}_{-0.1} \times10^{14}$  & $3.5^{+0.2}_{-1.1} \times10^{-2}$  \\
C$_2$H          & $J$=7/2-5/2 $F$=3-2   & 25.1      & -4.13 & 7    & 37.14     & 72.91     & 144.71    & $3.4^{+2.0}_{-1.5} \times10^{14}$  & $7.4^{+1.7}_{-1.8} \times10^{-2}$  \\
C$_2$H          & $J$=5/2-3/2 $F$=3-2   & 25.2      & -4.15 & 7    & 37.14     & 72.91     & 144.71    & $1.8^{+0.3}_{-0.1} \times10^{14}$  & $3.8^{+2.2}_{-0.1} \times10^{-2}$  \\
c-C$_3$H$_2$    & 3$_{21}$-2$_{12}$ & 18.2      & -4.23 & 21    & 201.84    & 566.86    & 1597.89   & $5.7^{+4.0}_{-1.6} \times10^{12}$  & ...$^b$\\
H$^{13}$CN      & 3-2               & 24.9      & -3.11 & 21    & 28.17     & 55.31     & 109.62    & $5.4^{+0.4}_{-1.3} \times10^{12}$  & $6.9^{+3.0}_{-3.0} \times10^{-1}$ \\
HC$^{15}$N      & 3-2               & 24.8      & -3.12 & 7     & 9.42      & 18.50     & 36.66     & $1.1^{+0.2}_{-0.1} \times10^{12}$  & $1.4^{+1.2}_{-0.5} \times10^{-1}$ \\
DCN             & 3-2               & 20.9      & -3.34 & 21    & 33.39     & 65.75     & 130.51    & $3.7^{+0.5}_{-0.3} \times10^{12}$  & $3.6^{+3.7}_{-1.4} \times10^{-1}$ \\
H$^{13}$CO$^+$  & 3-2              & 25.0      & -2.88 & 7     & 9.35      & 18.35     & 36.37     & $2.2^{+0.3}_{-0.1} \times10^{12}$  & $4.9^{+4.7}_{-1.9} \times10^{-1}$  \\
CS              & 5-4               & 35.3      & -3.53 & 11    & 16.28     & 32.24     & 64.15     & $1.0^{+2.6}_{-0.6} \times10^{13}$  & $5.0^{+4.8}_{-1.8} \times10^{-1}$ \\
SO              & 7$_6$-6$_5$               & 47.5      & -2.11 & 15 	& 38.88 	& 90.35 	& 197.51     & $<1.5\times10^{10}$  & ...$^c$ \\
SO              & 6$_6$-5$_5$               & 56.4      & -2.21  & 13 	& 38.88 	& 90.35 	& 197.51     & $<7.1\times10^{10}$  & ...$^c$
\enddata
\tablenotetext{}{$^a$ Peak column densities and optical depths are derived for an excitation temperature of 30\,K, with the exception of H$_2$CO where $T_{\rm ex}$ has been computed self-consistently. The error bars include the statistical uncertainties from the posterior distributions, a 10\% flux calibration uncertainty, and for all lines but H$_2$CO the systematic uncertainty in the excitation temperature. $^b$ No radial profile was derived for c-C$_3$H$_2$. The column density represents the disk averaged value, computed over a disk radius of $3\arcsec$ using the flux from Tab.~\ref{tab:fluxes}. $^c$ The $3\sigma$ upper limit on the line flux has been used to derive an upper limit on the disk averaged column density, using the same approach as for c-C$_3$H$_2$.}
\end{deluxetable*}

\subsection{Deuteration and N fractionation}
\label{sec:fractionation}

\begin{figure*}
\begin{center}
\begin{tabular}{c}
\includegraphics[width=0.45\textwidth]{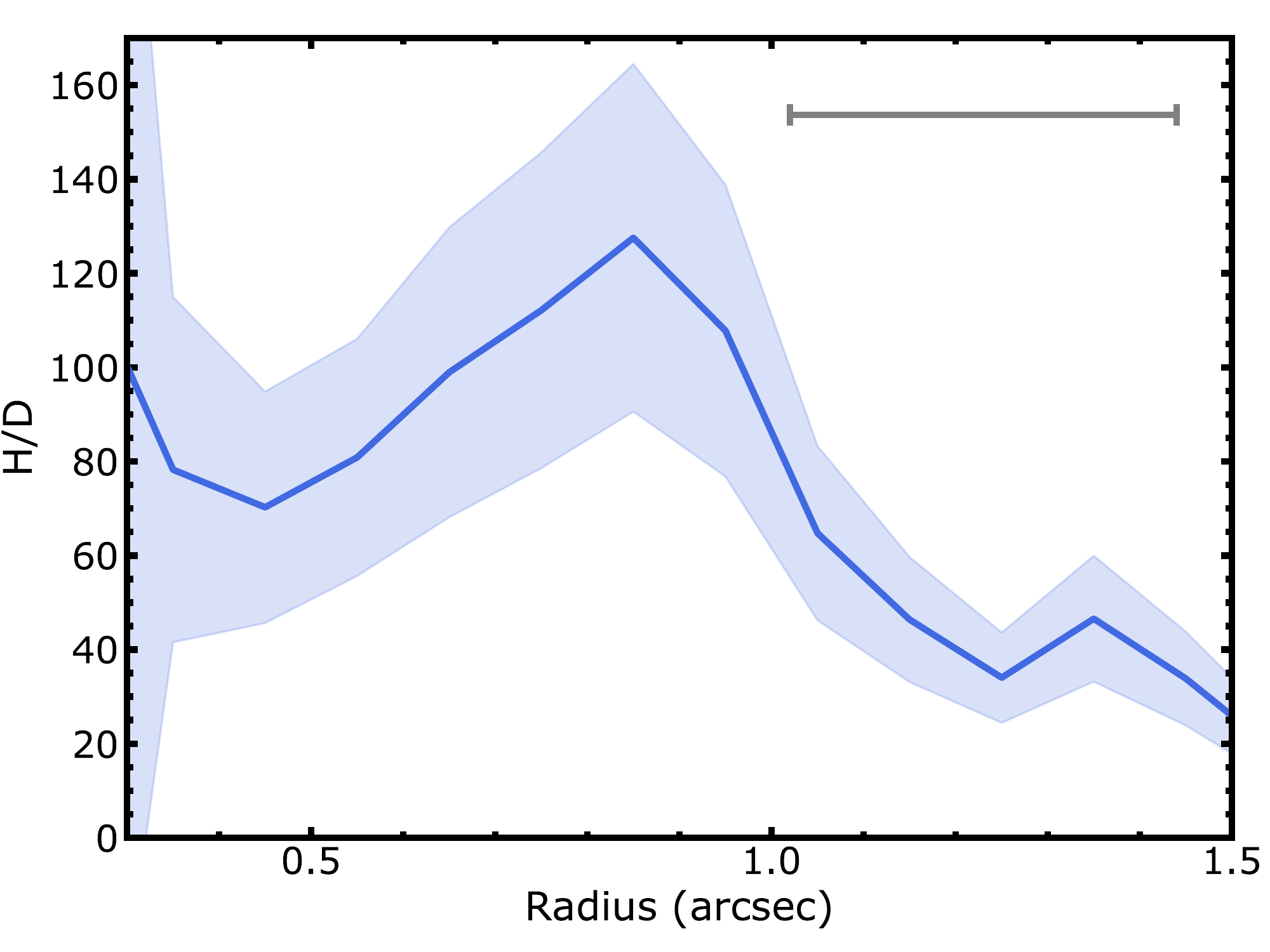}
\includegraphics[width=0.45\textwidth]{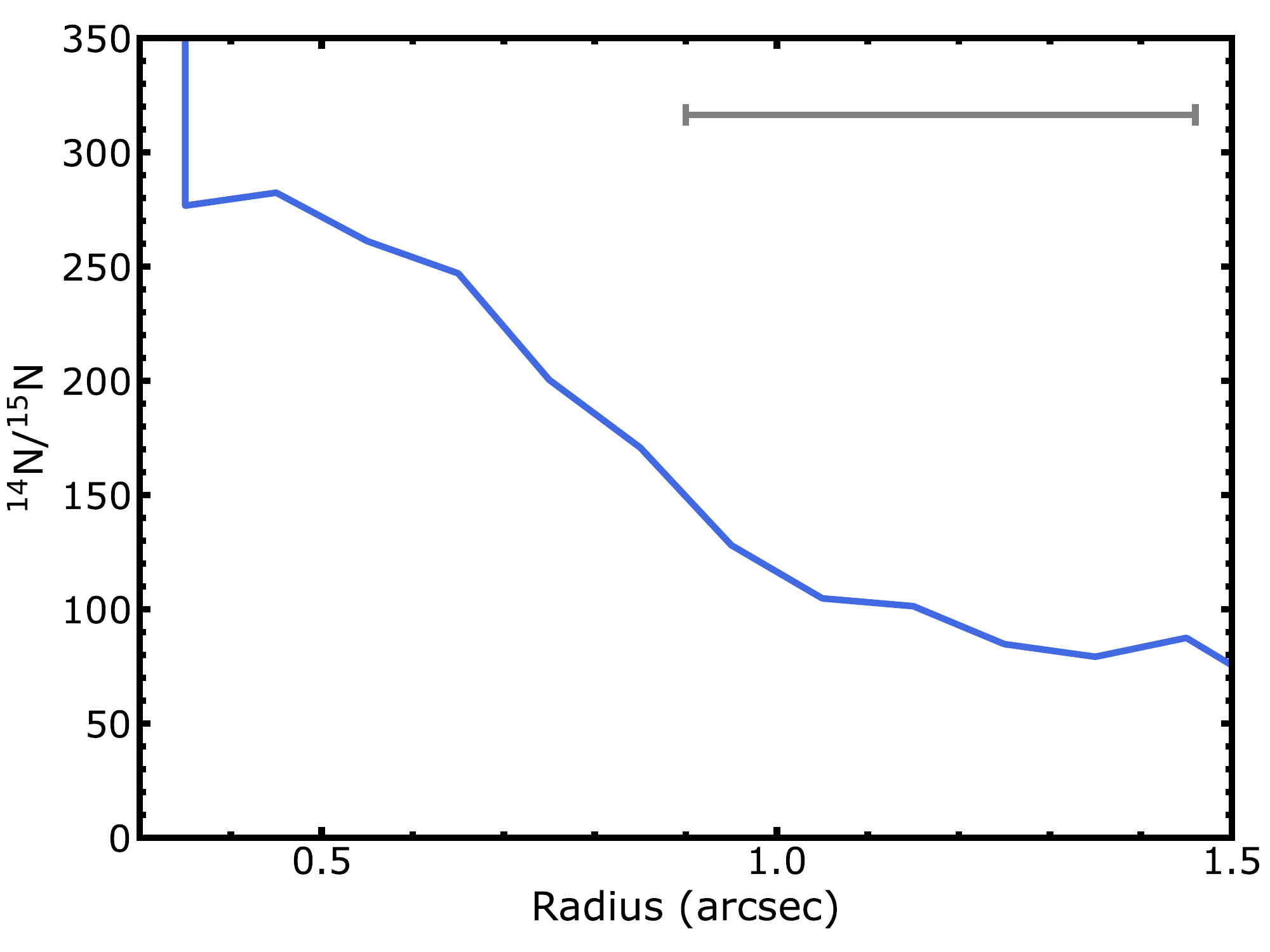}\\
\end{tabular}
\end{center}
\caption{Left: H/D ratio calculated as $N$(HCN)/$N$(DCN), where $N$(HCN) has been computed from the H$^{13}$CN column density. Right: $^{14}$N/$^{15}$N ratio calculated as $N$(HCN)/$N$(HC$^{15}$N). The grey lines at the top right show the beam major axis, after smoothing the images used in the ratios to the same beam. The ribbons indicate $1\sigma$ uncertainties, which include the standard deviation and rms divided by the number of beams along the azimuthal ring (left), the uncertainty on the $^{12}$C/$^{13}$C ratio, on the excitation temperature, and for the $N$(HCN)/$N$(DCN) ratio on the absolute flux due to calibration uncertainties.}
\label{fig:fractionation}
\end{figure*}

The hydrogen cyanide isotopologues can be used to infer the level of deuteration and nitrogen fractionation in PDS~70. The H$^{13}$CN/DCN and H$^{13}$CN/HC$^{15}$N ratios have been already used in the literature to extract disk-average deuteration and fractionation values in a sample of disks comprising about 10 sources \citep[e.g.][]{huang2017,guzman2017,bergner2019,bergner2020}. We follow the same procedure used in \citet{bergner2020} to extract the column density ratio between the three isotopologues. In order to derive HCN column densities, a $^{12}$C/$^{13}$C ratio of 68$\pm15$ is assumed \citep{milam2005}. While the carbon fractionation is still not well established in the majority of disks, \citet{hily-blant2019} derived a uniform carbon fractionation across the TW Hya disk using HCN and H$^{13}$CN lines, with a value compatible with the one we assume. The column density ratio between two isotopologues is derived using Eq.~\ref{eq:column}, and taking the ratio between two different isotopologues. As noticed by \citet{guzman2017}, the similar upper state energy and $Q/g_{\rm u}$ ratios of the three isotopologues leads to a column density ratio that is very similar to a flux ratio. We assume that the emission from the three isotopologues is at the same excitation temperature, which we fix at 30\,K. As in Section~\ref{sec:column}, the temperature is varied between $20-50\,$K in the computation of the uncertainties on the ratio. The systematics on the excitation temperature leads to relatively small uncertainties in the ratio, between 7\% for the H/D ratio, and 0.1\% for the $^{14}$N/$^{15}$N ratio (where we assumed the molecules to be at the same temperature in the uncertainty calculation). Other sources of uncertainty are included in the error estimate (flux rms, $^{12}$C/$^{13}$C, absolute flux calibration), and in all cases dominate the uncertainties over the excitation temperature systematics. The absolute flux calibration uncertainty is not included for H$^{13}$CN/HC$^{15}$N, since the two transitions are very close in frequency in the spectral setup (we neglect any uncertainty in the bandpass calibration). The derived disk averaged H/D and $^{14}$N/$^{15}$N in HCN isotopologues are $60\pm17$ and $111\pm27$, respectively, where a radius of $3\arcsec$ has been used to extract the disk averaged values.

The two values show a significant D and $^{15}$N enrichment in HCN isotopologues relative to proto-solar elemental ratios. In Jupiter atmosphere, D/H $\sim 2.5\times10^{-5}$ from the HD/H$_2$ ratio \citep{Linksy_ea_2006}, while solar winds indicate a ratio $^{14}$N/$^{15}$N$\sim441\pm5$ \citep[][]{marty2011}. Fractionation levels of HCN isotopologues in both nitrogen and hydrogen show consistently lower values in protoplanetary disks. In particular, \citet{huang2017} showed an H/D ranging between $\sim15 - 200$ in a sample of six disks, based on DCN and H$^{13}$CN lines. \citet{guzman2017} obtained a $^{14}$N/$^{15}$N between $\sim80-160$ in the same sample by analyzing H$^{13}$CN and HC$^{15}$H transitions. PDS~70 falls well within these distributions. The nitrogen fractionation of HCN in PDS~70 is in line with values observed in solar systems bodies, e.g. comets showing HCN/HC$^{15}$N$\approx 100-250$ \citep{mumma2011}. Such low ratio are comparable with HCN isotopologues observations of pre-stellar cores \citep[e.g.,][]{hily-blant2013a}, even though new observations suggest that there may be a large variation between different cores \citep{Magalhaes2018}, in particular in the high mass regime \citep{colzi_ea_2018}. Thus the high nitrogen fractionation inferred in PDS~70 may be inherited from its progenitor core.

We derive the spatial dependence of H/D and $^{14}$N/$^{15}$N using column densities of the three HCN isotopologues. Purely inherited material would show uniform fractionation and deuteration across the radial extent of the disk \citep[e.g.,][]{guzman2017,hily-blant2019}, while radial gradients would be indicative of continued chemical processing. Radial profiles of the column densities are computed as described in Section~\ref{sec:column} from integrated intensity maps created with matching synthesized beams. In particular, the H$^{13}$CN and DCN lines are compared with maps obtained with a robust=0 parameter as in Section~\ref{sec:moment0}, and then smoothed to a common circular beam of $0\farcs42$ with the CASA \texttt{imsmooth} task. The H$^{13}$CN and HC$^{15}$N lines are instead compared from images with natural weighting, which lead to the same beam. An excitation temperature of 30\,K is assumed, and uncertainties on $T_{\rm ex}$ are considered in the error propagation. Other sources of uncertainties are considered, including the uncertainty on $^{12}$C/$^{13}$C, as for the disk average values. The column density ratios are shown in Fig.~\ref{fig:fractionation}. The figures include the derived radial profiles of $N$(HCN)/$N$(DCN) and $N$(HCN)/$N$(HC$^{15}$N), with $N$(HCN) being $68\times N$(H$^{13}$CN). The analysis is limited to $r\lesssim 1\farcs8$ due to the low snr at larger radii.

Both D/H and $^{14}$N/$^{15}$N profiles show some radial dependence, with enhanced deuteration and nitrogen fractionation in outer regions of the disk. To our knowledge, this is the first time that a radial profile of the deuteration of HCN has been measured in a protoplanetary disk. \citet{huang2017} suggest that LkCa~15 may present a similar behavior, with enhanced deuteration in the outer regions of the disk, outside the mm ring of this transition disk. Since the formation of DCN is mediated by CH$_2$D$^+$ \citep[e.g.,][]{millar_ea_1989,turner_2001}, a high level of deuterium fractionation may indicate favorable deuterium exchange between HD and C$_2$H$_2^+$ at temperatures $<80$\,K outside the pebble disk at lower temperatures (see Section~\ref{sec:radial_profiles}). Finally, the HCN deuteration profile indicates that the dust ring at $0\farcs7-0\farcs8$ may affect the deuteration process, with the $N$(H$^{13}$CN)/$N$(DCN) profile showing a peak co-located with the dust mm thermal emission. This coincides with the temperature increase in the H$_2$CO excitation temperature profile (see Fig.~\ref{fig:h2co}), thus directly tracing a less efficient deuterium fractionation at higher temperatures in the disk, as expected from chemical models. The correlation of the $N$(H$^{13}$CN)/$N$(DCN) peak with the sub-mm continuum ring highlights the direct impact that dust substructure can have on chemical processes in protoplanetary disks, including isotope exchange reactions.

The radial profile of the nitrogen fractionation is particularly surprising. \citet{guzman2017} and \citet{hily-blant2019} observed radially decreasing fractionation levels in HCN isotopologues in the V4046 Sgr and TW Hya disks, respectively. \citet{hily-blant2019} observed tentative evidence of increasing fractionation levels in the very outer disk, but the low snr of the observations precluded a robust conclusion. A radially decreasing level of fractionation can be explained by enhanced fractionation in the inner regions of the disk due to isotope selective photo-dissociation of the self-shielding molecule N$_2$ \citep[][]{heays2014,visser_ea_2018}. Higher UV fluxes in the inner regions can selectively photodissociate N$^{15}$N, with the $^{15}$N atom free to enter gas-phase chemical reactions and lead to formation of HC$^{15}$N. However, the radially decreasing fractionation in PDS~70 suggests that in the outer disk the main fractionation mechanism is low temperature isotope exchange reactions \citep[e.g.,][]{terzieva_ea_2000,roueff_ea_2015,wirstrom_ea_2018}, which dominates over the isotope selective photodissociation pathway. It is not clear why the behavior in PDS~70 is so different from the TW Hya and V4046 Sgr disks. The most likely possibility is that the inner cavity and the bright dust wall at the cavity edge significantly affects the temperature and radiation structure across the whole disk, thus leading to favorable low temperature reactions outside the cavity wall. The radial polarized intensity profile in $J$-band \citep[see Figure~2 in][and Section~\ref{sec:comparison_peaks} in this paper]{keppler18} shows a pronounced peak at $\sim0\farcs48$ from the central star, suggesting that the outer regions may be shadowed by the warm wall of material at that cavity edge, as predicted by theoretical models \citep{isella_and_turner_2018}.

As specified earlier, the radial profiles assume a uniform $N$(HCN)/$N$(H$^{13}$CN) ratio. However, \citet{visser_ea_2018} noted that from thermochemical models high $N$(HCN)/$N$(H$^{13}$CN) are found at large radii ($r>150$\,au in their model). We thus caution that our conclusions about deuteration and nitrogen fractionation may be affected by the carbon fractionation profile, were this not uniform \citep[e.g.,][]{colzi_ea_2020}. Observations of the main HCN isotopologue are needed to break this degeneracy \citep[e.g.,][]{hily-blant2019}.

Our observations indicate that organic material in the disk of PDS~70 possesses high nitrogen and hydrogen fractionation levels. Thus, organic material accreted by the two planets will possess high fractionation levels, which may alter the isotope ratios of the planetary atmospheres.

\section{Discussion}
\label{sec:discussion}

\subsection{C/O ratio}

The elemental carbon-to-oxygen (C/O) ratio in the atmosphere of giant planets has been proposed as a probe of the accretion history of massive protoplanets during their co-evolution with their host protoplanetary disk \citep[e.g.,][for some recent reviews]{Putritz_ea_2018,Lammer_ea_2018,madhu2019,oberg_and_bergin_2020,van_dishoeck_ea_2020}. While molecular phase transitions at specific temperatures across the disk can affect the elemental ratio of gas phase material \citep[e.g.,][]{oberg2011}, additional chemical evolution within the disk can alter the C/O ratio and thus the chemical outcome of the planet formation process \citep[e.g.,][]{eistrup_ea_2016,eistrup_ea_2018,cridland_ea_2019a,Cridland_2020}. Probing the C/O ratio in disks is thus key in accessing whether the combination of both chemical and physical evolution of gas and dust in disks has affected the elemental abundances available in the disk gas phase material \citep[e.g.,][]{kama_ea_2016,zhang_ea_2019,krijt_ea_2020}. 

To determine the C/O ratio in the vertical layers of disks that can be probed by ALMA observations, emission lines of small hydrocarbons are extremely valuable. In particular, the abundance of C$_2$H increases very rapidly as soon as C/O$>$1, where free C atoms are not locked up in CO, thereby quenching complex organic chemistry. For this reason, the large C$_2$H fluxes of rotational transitions have been used to infer C/O ratios above the solar 0.54 value (in some cases above unity) in a few protoplanetary disks \citep[e.g.,][]{bergin2016,Cleeves2018,miotello_ea_2019}.

Comparing the C$_2$H luminosities of disks from the surveys by \citet{bergner2019,miotello_ea_2019}, PDS~70 is on the high flux range of the Lupus survey \citep[accounting for source distances, see Fig.~ 4 in][]{miotello_ea_2019}, with a total flux of the $J$=$\frac{7}{2}$-$\frac{5}{2}$ transition of $\sim1327\pm18\,$mJy\,km\,s$^{-1}$ (see Table~\ref{tab:fluxes}). In the thermo-chemical models by \citet{miotello_ea_2019}, such a high flux cannot be obtained with a solar C/O ratio, and elemental ratios close to $\sim1.5$ are required. Similarly, \citet{fedele_ea_2020} proposed to normalize line fluxes to the $^{13}$CO $J$=2-1 line to ease comparisons between sources. By exploring part of the wide parameter space that disks can exhibit in their physical and stellar properties, they showed that C/O$>1$ is required when the ratio of the C$_2$H $N$=3-2 $J$=$\frac{7}{2}$-$\frac{5}{2}$ $F$=4-3 line to the $^{13}$CO $J$=2-1 line is $\gtrsim 0.2$. For PDS~70, this ratio ranges between 0.25-0.35 throughout the whole disk out to $r\sim2\arcsec$, indicating again a C/O ratio above unity for this particular disk. However, we note that all these models have been fine tuned for disks that do not exhibit a large cavity, as is the case for PDS~70. Accurate C/O ratio retrievals need a forward model of the PDS~70 disk, which will be the focus of future works.

\begin{figure*}
\begin{center}
\begin{tabular}{c}
\includegraphics[width=0.8\textwidth]{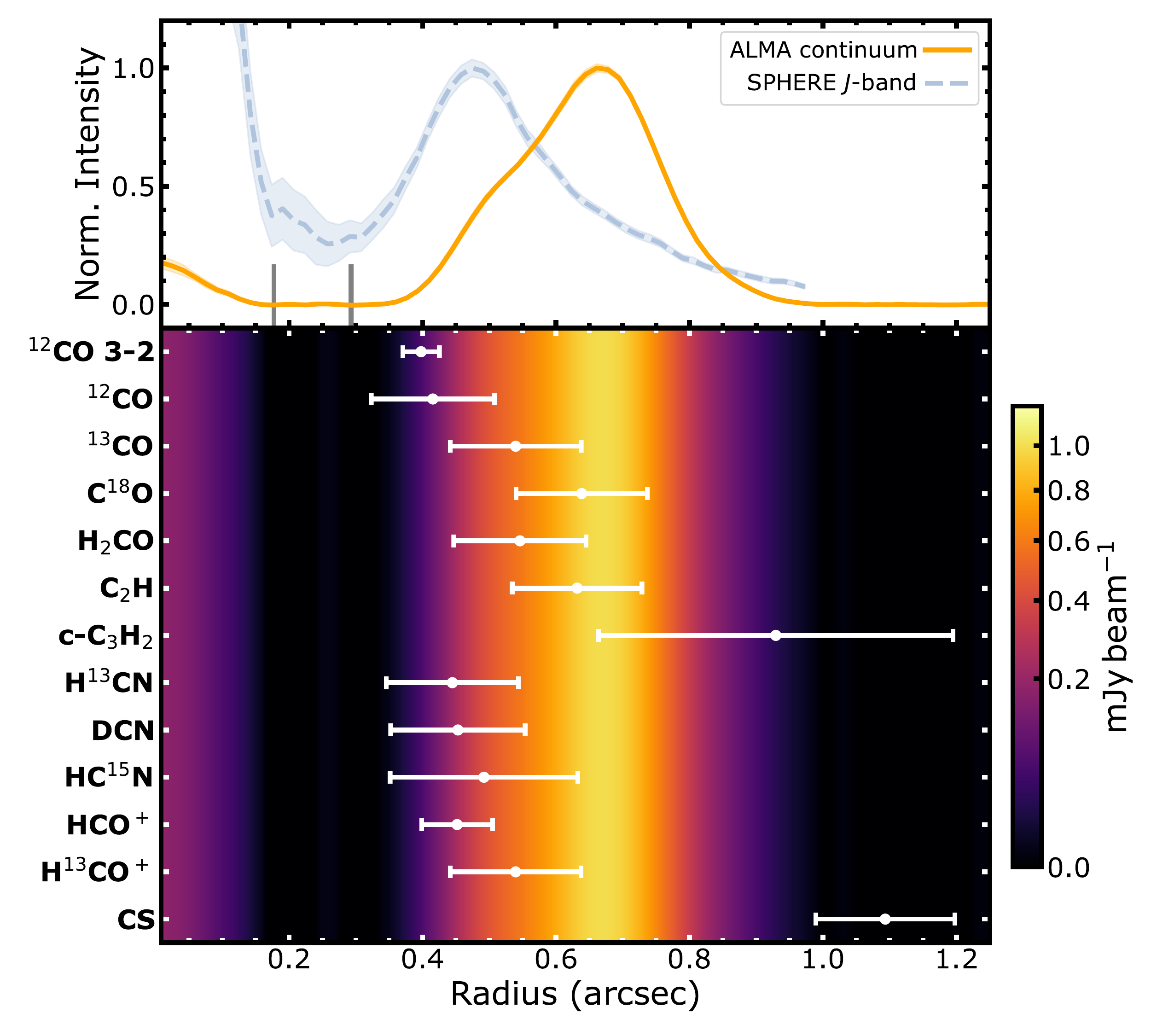}\\
\end{tabular}
\end{center}
\caption{Top panel: normalized 855$\,\mu$m continuum radial intensity profile \cite[data by][]{isella19}, and normalized $J$-band polarized intensity profile \citep[data by][]{keppler18}, with the two grey markers highlighting the orbital radius of PDS~70b and c \citep{wang_ea_2021}. Bottom panel: the ALMA continuum intensity profile is shown in color code in physical units, where the synthesized beam is $67\times50\,$mas. White dots indicate the radius of the integrated intensity profile peaks for all the detected species, listed on the $y$-axis. The C$_2$H represents the $J$=$\frac{7}{2}$-$\frac{5}{2}$ line including both detected hyperfine components; H$_2$CO is shown using the brighter 3$_{03}$-2$_{02}$ line.  The $1\sigma$ error bars represent one quarter of the major axis of the beam of the respective images used to extract the radial profiles, which are listed in Table~\ref{tab:images}.}
\label{fig:peak_radius}
\end{figure*}

An alternative way to qualitatively determine the C/O ratio in PDS~70 is by looking at the $N$(CS)/$N$(SO) ratio, as suggested by \citet{semenov_ea_2018} for the DM Tau disk \citep[see also][for a larger disk sample]{dutrey_ea_2011,guilloteau_ea_2016}. With a grid of thermo-chemical models, the authors show that the only models able to reproduce their observations of sulfur-bearing molecules (in particular CS and SO) is with a C/O ratio $\gtrsim1$. Their reference model produces a disk averaged column density ratio $N$(CS)/$N$(SO)$\sim0.03$, while observations indicate a ratio with a lower limit $>1$. Other chemical and physical parameters can affect the column density ratio, but not as significantly as the C/O ratio.

We apply the same qualitative argument to the PDS~70 disk. By using the $3\sigma$ upper limit on the SO column density, and the disk averaged CS column density of $1.8^{+0.4}_{0.2}\times10^{12}$\,cm$^{-2}$, we obtain a lower limit on the $N$(CS)/$N$(SO) ratio of $\sim109$. This value is $4000$ times higher than what \citet{semenov_ea_2018} predicted for DM Tau with a solar C/O ratio. A C/O ratio of $\sim1$ leads to a $N$(CS)/$N$(SO) ratio of $\sim10$ \cite{semenov_ea_2018}. Our lower limit therefore strongly suggests that the upper molecular layers of PDS~70 have a C/O ratio above unity, with significant amount of oxygen being sequestered in likely refractory form. Thermo-chemical models of PDS~70 are needed to extract a more quantitative conclusion.

In summary, a comparison of the PDS~70 inventory with  thermo-chemical models of protoplanetary disks from the literature suggests a disk-average C/O ratio $>1$ using different proxies, making this qualitative conclusion robust against different methods.

\subsection{Comparing mm Continuum, IR Scattered Light and Molecular Emission Profiles}
\label{sec:comparison_peaks}

In Section~\ref{sec:radial_profiles}, we have shown that all molecular transitions well detected in the ALMA data exhibit a ring-like morphology, with a high level of azimuthal symmetry. Molecular emission is observed to emit in ring-like morphologies in many disks and with different tracers \citep[e.g.,][]{oberg2015,Teague16,bergin2016}, with the emission morphology being primarily caused by chemical and excitation effects \citep[e.g.,][]{oberg2015,cazzoletti2018,long_ea_2021}, rather than the underlying dust density structure. However, gas structure can lead to ring-like morphologies in molecular transitions more sensitive to gas densities, such as CO isotopologues, as is clearly the case in extreme conditions as for transition disks \citep[e.g.,][]{vandermarel2016}. The presence of gaps and rings in dust is also expected to lead to molecular abundance variations along the radial coordinate from thermo-chemical models \citep[e.g.,][]{Facchini2018,rab2020,alarcon2020}. To assess how different is the molecular emission morphology across different lines in PDS 70, and whether these are related to the underlying dust density distribution as traced by mm continuum thermal emission, the radial peak of the integrated intensity profile is analyzed. From the radial profiles of Section~\ref{sec:radial_profiles}, the radius of the peak intensity is derived without any smoothing or fitting. An error of one quarter of the beam major axis is assigned to the peak location. The measured values are reported in Table~\ref{tab:images}. For the two $^{12}$CO lines and for HCO$^+$ only the off-center peak at $r>0.2\arcsec$ is analyzed.

The summary plot comparing the peak locations to the 855$\,\mu$m continuum emission presented in \citet{isella19} and the $J$-band polarized intensity by \citet{keppler18} is shown in Figure~\ref{fig:peak_radius}. The lines peak at different locations in the disk, with the large majority peaking within the radial maximum of the sub-mm continuum emission. The only two exceptions are CS and c-C$_3$H$_2$, peaking at $r\gtrsim0.9\arcsec$ outside the continuum ring, similarly to what is observed in other systems \citep[e.g.,][]{Teague16,cleeves_ea_2021}. This general behavior indicates that the underlying gas density structure significantly affects most of the molecular lines emission. At the same time, the variety of peak locations within the sub-mm cavity indicates how the flux of molecular lines is determined by a combination of factors (density, kinetic temperature, excitation temperature, optical depth, molecular abundance, etc.). While reproducing all these emission lines would require significant thermo-chemical modeling, some trends can be identified heuristically. The most optically thick lines, such as $^{12}$CO, peak at the innermost radii, indicating the high temperatures at the edge of the cavity wall. Rarer CO isotopologues, namely $^{13}$CO and C$^{18}$O, progressively peak at larger radii, with the lines better tracing bulk gas densities as they become optically thinner. Interestingly, the C$^{18}$O peak is co-located with the peak in dust continuum, likely indicating a density maximum leading to efficient dust trapping. The HCN isotopologues lines, which are found to have low optical depths (see Section~\ref{sec:fractionation}), peak at smaller radii than the optically thicker C$^{18}$O. This likely shows an excitation and abundance effect, with HCN being mostly abundant at the gas cavity wall, where the outer edge of the gap carved by the planet is directly illuminated by UV photons from the central star triggering the production of HCN via vibrationally excited H$_2$ \citep[e.g.,][]{visser_ea_2018, cazzoletti2018}. This is also indicated by their emission peaking at the same radial range of the peak in the IR scattered light image, which traces the cavity edge in small grains. Similarly, HCO$^+$ and H$^{13}$CO$^+$ peak at a similar radial range, likely showing a high abundance of their precursor H$_3^+$ where X-ray photons from the central star ionizing H$_2$ impinge onto the cavity wall \citep[e.g.,][]{cleeves_ea_2014,cleeves_ea_2017}. This indicates that group of molecules for which the emission significantly determined by the illumination pattern correlate better with IR scattered light images, highlighting the illumination pattern onto the disk, than with mm continuum images, which better highlight the bulk dust density structure.

In summary, the radial profiles of the different molecules show that the gas and dust densities, shaped by the massive planets carving a large gap visible in $^{12}$CO, affect the molecular emission. The large cavity of PDS~70 is an extreme case of a gapped structure in the surface density profile, and it can be used to constrain thermo-chemical models focusing on the effect of gaps and rings in affecting chemical abundances and molecular emission in protoplanetary disks \citep[e.g.,][]{Facchini2018,rab2020,alarcon2020}. In PDS~70, we have shown that different molecules peak at a variety of radial scales, with separate group of molecules being sensitive to different physico-chemical factors. These observations will be at the foundation of upcoming targeted thermo-chemical models of PDS~70 to determine the physico-chemical interplay between dust and gas structure and molecular emission in transition disks.

\section{Conclusions}
\label{sec:conclusions}

In this paper, we have presented the chemical inventory of the planet hosting disk PDS~70. 
The conclusions can be summarized with the following:

\begin{enumerate}
    \item Achieving a typical sensitivity of $\sim1\,$mJy\,beam$^{-1}$ in a 1\,km\,s$^{-1}$ channel with an angular resolution of $\sim0\farcs4-0\farcs5$, we detect 16 transitions from 12 molecular species: $^{12}$CO, $^{13}$CO, C$^{18}$O, H$_2$CO, C$_2$H, c-C$_3$H$_2$, H$^{13}$CN, HC$^{15}$N, DCN, HCO$^+$, H$^{13}$CO$^+$ and CS. No emission from CH$_3$OH or SO is detected.
    \item Integrated intensity maps and radial profiles of the integrated intensities show that the molecular emission originate primarily from a bright ring outside the orbit of the two detected planets. Only the two $^{12}$CO and the HCO$^+$ lines are associated with emission from gas within the cavity carved by planets b and c.
    \item Molecular lines peak at different radial locations, highlighting the complex interplay of temperature, density and stellar irradiation in determining molecular abundances. The molecular structure indicates a chemically active layer at the edge of the cavity wall of the transition disk.
    \item Column densities and optical depths are derived for all molecules with optically thin transitions. The excitation temperature is computed self-consistently for H$_2$CO, where two transitions are detected. For all other molecules a range of excitation temperatures is considered to estimate the uncertainty on the column density.
    \item Deuteration and nitrogen fractionation profiles are derived from the hydrogen-cyanide isotopologues. Both show a high fractionation level, with fractionation increasing at larger radii, possibly suggesting a cold deuteration and fractionation pathway outside the mm-emitting ring.
    \item The high C$_2$H/$^{13}$CO flux ratio and a high lower limit on the $N$(CS)/$N$(SO) column density ratio indicate a disk average C/O ratio above unity. This suggests that the actively accreting protoplanets are likely to being enriched with high C/O ratio gas material.
    \item The emission of some molecular transitions, such as C$^{18}$O $J$=2-1, correlates well with the (sub-)mm continuum emission profile, thus indicating that they are good tracers for the bulk disk density profile. Other group of molecules, such as HCN and HCO$^+$ isotopologues, peak in the same radial range of the IR scattered light intensity profile, highlighting the effect of direct illumination on the chemistry at the cavity edge.  
\end{enumerate}

PDS~70 provides a unique opportunity to witness how interactions between newly formed planets and their parental disk can sculpt the gas and dust distributions and hence influencing the chemical complexity of the atmosphere-building material that is actively accreted onto the planets. This work presents the first chemical inventory of the PDS~70 system, revealing a rich chemistry where radial variations in the line emission suggest a dynamical chemical structure which reacts to the planet-driven perturbations in the underlying physical structure. Upcoming high resolution data, probing spatial scales of $\sim$10~au, will facilitate a comprehensive study of the chemical complexity at key locations in the disk, such as the highly irradiated cavity edge and along fast radial flows delivering the atmosphere-building material to the giant planets. Together with present and future projects trying to determine the chemical properties and C/O ratio in the atmospheres of PDS~70b and c \citep[e.g.,][]{wang_ea_2021}, such work will provide the most direct link between the formation environment of an exoplanet and the elemental ratios found in its atmosphere, a key science case of current and upcoming facilities.

\acknowledgments{We thank the referee for their comments on the manuscript. We are thankful to Ewine F. van Dishoeck for early suggestions that helped defining this paper. We appreciate useful conversations with Ted Bergin, Dmitry Semenov, Romane Le Gal and Teresa Paneque. We thank Mark Gurwell for performing the re-calibration of the archival SMA data. This paper makes use of the following ALMA data:\\
ADS/JAO.ALMA\#2015.1.00888.S, \\ ADS/JAO.ALMA\#2017.A.00006.S, \\ ADS/JAO.ALMA\#2019.1.01619.S. ALMA is a partnership of ESO (representing its member states), NSF (USA), and NINS (Japan), together with NRC (Canada),  NSC and ASIAA (Taiwan), and KASI (Republic of Korea), in cooperation with the Republic of Chile. The Joint ALMA Observatory is operated by ESO, AUI/NRAO, and NAOJ. SF acknowledges an ESO Fellowship. This project has received funding from the European Union’s Horizon 2020 research and innovation program under the Marie Sklodowska-Curie grant agreement No 823823 (Dustbusters RISE project). JB acknowledges support by NASA through the NASA Hubble Fellowship grant \#HST-HF2-51427.001-A awarded  by  the  Space  Telescope  Science  Institute,  which  is  operated  by  the  Association  of  Universities  for  Research  in  Astronomy, Incorporated, under NASA contract NAS5-26555.}

\facility{ALMA}

\software{\texttt{GoFish} \citep{gofish}, \texttt{VISIBLE} \citep{VISIBLE}, \texttt{GALARIO} \citep{galario}, \texttt{CASA} \citep{casa}, \texttt{Matplotlib} \citep{matplotlib}, \texttt{numpy} \citep{numpy}, \texttt{emcee} \citep{Foreman-Mackey+13}.}

\appendix

\section{Channel maps of weak lines}
\label{sec:channel_map}

\begin{figure*}
\begin{center}
\begin{tabular}{c}
\includegraphics[width=\textwidth]{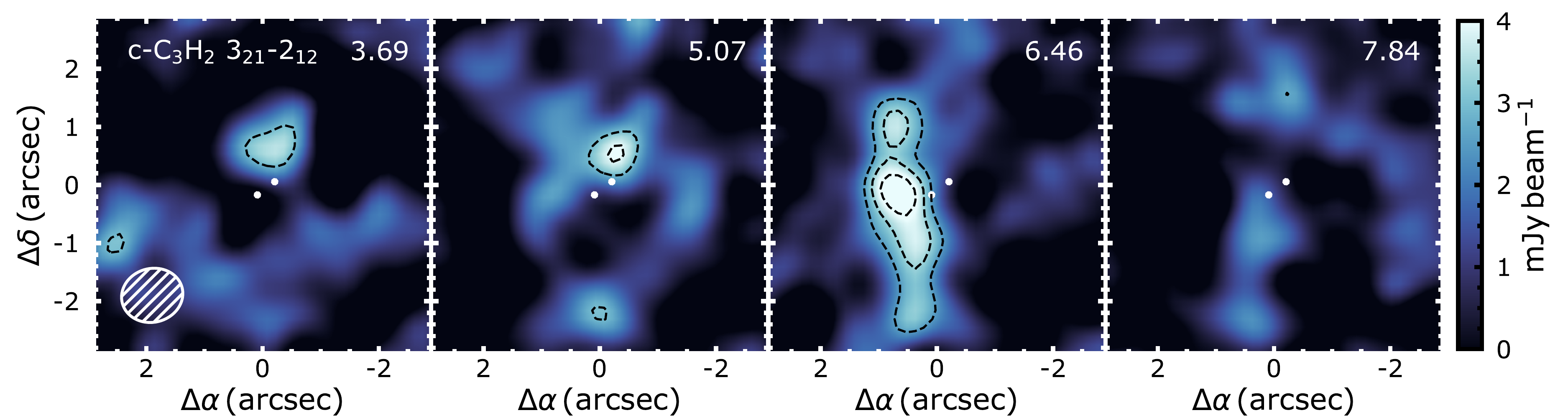}\\
\includegraphics[width=\textwidth]{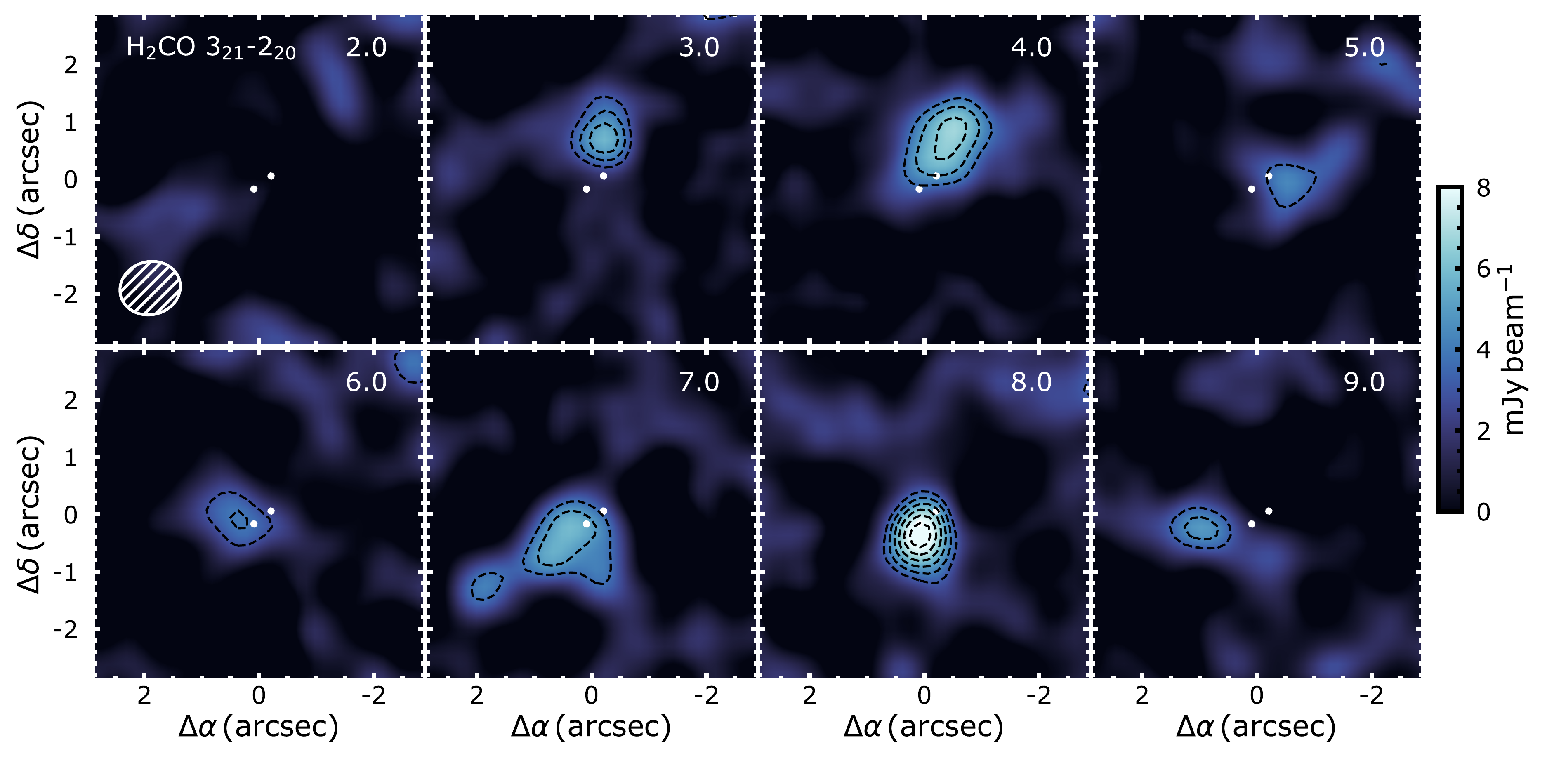}
\end{tabular}
\end{center}
\caption{Channel maps of the c-C$_3$H$_2$ 3$_{21}$-2$_{12}$ (top) and H$_2$CO 3$_{21}$-2$_{20}$ (bottom) transitions, imaged with natural weighting and an additional $uv$-tapering of $0\farcs7$. The channel velocity is reported in the top right corner of the panels in units of km\,s$^{-1}$. Dashed contours indicate the [3, 4, 5, 6, 7, 8]$\sigma$ levels. The two white dots indicate the location of planets b and c.}
\label{fig:channel_maps}
\end{figure*}

The channel maps of the c-C$_3$H$_2$ 3$_{21}$-2$_{12}$ and H$_2$CO 3$_{21}$-2$_{20}$ transitions are reported in Fig.~\ref{fig:channel_maps}. The images were produced with natural weighting and an additional $uv$-tapering of $0\farcs7$. The properties of the images are listed in Sec.~\ref{sec:fluxes}.

\section{Radial profiles with GoFish}
\label{app:radial_profiles}

\begin{figure*}
\begin{center}
\begin{tabular}{c}
\includegraphics[width=0.95\textwidth]{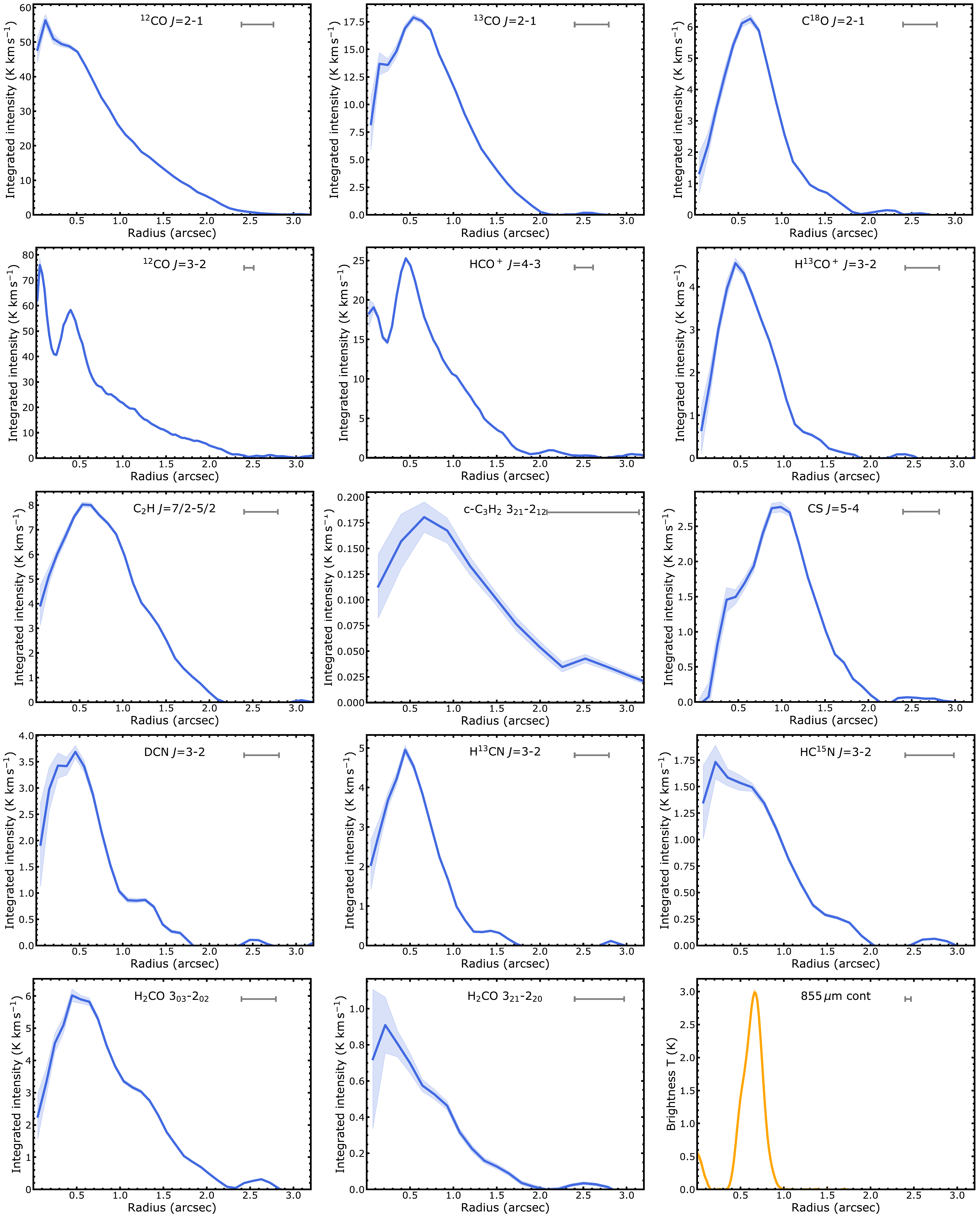}\\
\end{tabular}
\end{center}
\caption{Azimuthally averaged radial profiles of the 14 detected lines, and of the 885$\,\mu$m continuum by \citet{isella19} in the bottom right in comparison. The profiles have been obtained with the \texttt{GoFish} package (see Section~\ref{sec:radial_profiles}). The imaging parameters of the respective channel maps are listed in Table~\ref{tab:images}. The ribbon shows the $1\sigma$ rms, excluding the uncertainty from absolute flux calibration. The grey line at the top right of all panels shows the beam major axis. The brightness temperature conversion was done under the Rayleigh-Jeans approximation.}
\label{fig:radial_profiles_gofish}
\end{figure*}

In Section~\ref{sec:radial_profiles}, radial profiles of the detected emission lines are computed by azimuthally averaging the integrated intensity maps shown in Fig. ~\ref{fig:moment0}. A different method to produce the intensity profiles has also been tested. The methodology is the same as described in Section~\ref{sec:fluxes} by the \texttt{GoFish} package, by deprojecting, shifting and azimuthally averaging the spectra along radial annuli, as shown in Figure~\ref{fig:teardrop}. The $^{12}$CO $J$=3-2 and HCO$^+$ lines are deprojected assuming the vertical component of the emission, with the same parameters derived by \citet{keppler19}, and used to create the maps in the cleaning process (see parameters in Section~\ref{sec:reduction}). The 2D spectra are then collapsed along the velocity axis in order to obtain the radial profile of the integrated intensity. The integral in velocity is performed on the following intervals: $-1$-12\,km\,s$^{-1}$ for the $^{12}$CO lines; 0-11\,km\,s$^{-1}$ for the HCO$^{+}$ line; 0-12\,km\,s$^{-1}$ for the C$_2$H $J$=$\frac{7}{2}$-$\frac{5}{2}$ line; 1-10\,km\,s$^{-1}$ for all the other lines. The profiles obtained with this method are shown in Fig.~\ref{fig:radial_profiles_gofish}, to be compared against the ones from the azimuthal average of the integrated intensity maps shown in Fig.~\ref{fig:radial_profiles}. For radii larger than the beam major axis, the agreement between the two methods is remarkable, whereas the inner regions of the stacking method leads to an underestimate of the integrated intensity due to beam and velocity smearing reducing the average intensity. Since most of the analysis in the paper focuses on the inner two-three beams, radial profiles from integrated intensity maps were preferred.

\bibliography{bibliography}{}
\bibliographystyle{aasjournal}

\end{document}